\documentclass[aip, amsmath, amssymb, preprint, floatfix]{revtex4-1}
\usepackage{graphicx}
\usepackage{dcolumn}
\usepackage{bm}
\usepackage[utf8]{inputenc}
\usepackage[T1]{fontenc}
\usepackage{mathptmx}
\usepackage{color,soul}

\usepackage{verbatim}

\newcommand{\ts}[1]{\textsubscript{#1}}

\begin{document}

\title{Enhancing magneto-optic effects in two-dimensional magnets by thin-film interference} 

\author{F. Hendriks}
\email[]{f.hendriks@rug.nl}
\affiliation{Zernike Institute for Advanced Materials, University of Groningen, The Netherlands}
\author{M.H.D. Guimarães}
\email[]{m.h.guimaraes@rug.nl}
\affiliation{Zernike Institute for Advanced Materials, University of Groningen, The Netherlands}

\date{\today}

\begin{abstract}
The magneto-optic Kerr effect is a powerful tool for measuring magnetism in thin films at microscopic scales, as was recently demonstrated by the major role it played in the discovery of two-dimensional (2D) ferromagnetism in monolayer CrI$_3$ and Cr$_2$Ge$_2$Te$_6$.
These 2D magnets are often stacked with other 2D materials in van der Waals heterostructures on a SiO$_2$/Si substrate, giving rise to thin-film interference.
This can strongly affect magneto-optical measurements, but is often not taken into account in experiments.
Here, we show that thin-film interference can be used to engineer the magneto-optical signals of 2D magnetic materials and optimize them for a given experiment or setup.
Using the transfer matrix method, we analyze the magneto-optical signals from realistic systems composed of van der Waals heterostructures on SiO$_2$/Si substrates, using CrI$_3$ as a prototypical 2D magnet, and hexagonal boron nitride (hBN) to encapsulate this air-sensitive layer.
We observe a strong modulation of the Kerr rotation and ellipticity, reaching several tens to hundreds of milliradians, as a function of the illumination wavelength, and the thickness of the SiO$_2$ and layers composing the van der Waals heterostructure.
Similar results are obtained in heterostructures composed by other 2D magnets, such as CrCl$_3$, CrBr$_3$ and Cr$_2$Ge$_2$Te$_6$.
Designing samples for the optimal trade-off between magnitude of the magneto-optical signals and intensity of the reflected light should result in a higher sensitivity and shorter measurement times.
Therefore, we expect that careful sample engineering, taking into account thin-film interference effects, will further the knowledge of magnetization in low-dimensional structures.
\end{abstract}


\maketitle

Magneto-optical effects, such as the Kerr and Faraday effect, are key to unveiling the magnetic structure and spin behavior of low-dimensional systems. \cite{Petit2008, Rogez2009, McCord2015, Jiang2018, Gibertini2019}
In these effects, a change of the reflected or transmitted light intensity and polarization is (often linearly) related to the change of magnetization of the illuminated area.
When used in combination with microscopy techniques, magneto-optical signals can be used to image the magnetization of systems at the sub-micrometer scale, \cite{Dickson2005, Savoini2011, Lange2017} and when combined with ultrafast lasers, they give access to the magnetization dynamics at femtosecond timescales. \cite{Beaurepaire1996, Kirilyuk2010, Koopmans2010, Wu2010, Zhang2020}
The magneto-optic Kerr effect (MOKE) was instrumental for the discovery of two-dimensional (2D) ferromagnetism in monolayer CrI\ts{3} and Cr\ts{2}Ge\ts{2}Te\ts{6}. \cite{Huang2017, Gong2017}
Due to its non-destructive nature and easy implementation, MOKE and related magneto-optic effects, such as reflected magnetic circular dichroism, are one of the standard tools for the magnetic characterization of 2D van der Waals magnets. \cite{Huang2017, Gong2017, Bonilla2018, Fei2018, Gibertini2019}
For those measurements, 2D magnets are often stacked with other van der Waals materials on a substrate, such as hexagonal boron nitride (hBN) on SiO\ts{2}/Si substrates.
These layered systems can display strong thin-film interference effects which in turn affect their magneto-optical response.
At the start of the 2D materials revolution, it was discovered that exploiting these interference effects allowed for optical identification of graphene flakes, \cite{Novoselov2004, Abergel2007, Blake2007} providing a way for easily identifying graphene mono- or few-layers.
Later, the same techniques were used for identifying thin layers of other van der Waals meterials, such as transition metal dichalcogenides. \cite{Castellanos-Gomez2010, Late2012, Castellanos-Gomez2013, Li2013}
Also, the effects of thin-film interference on magneto-optical signals, and how to use these effects to enhance them, have been studied extensively in the context of metallic thin-films, \cite{Berreman1972, Yeh1979, Mansuripur1982, Zak1990, Cantwell2006, Sumi2018} oriented moleculecular films \cite{Brauer2009}, ellipsometry \cite{HUMLICEK20053, SCHUBERT2005637, Fujiwara2007}, and many other fields.
However, thin-film interference effects are often not taken fully into account for the magneto-optical experiments on van der Waals magnets. \cite{Huang2018, Wang2018a}
This could lead to a suboptimal signal-to-noise ratio, resulting in a lower sensitivity and / or longer measurement times. Therefore, it becomes more difficult and more time-consuming to measure small changes in magnetization of 2D magnets, caused by for example chiral spin textures in a homogeneously magnetized lattice, and to measure under low-light conditions to avoid sample degradation.
While some works do take into account the effect of the oxide substrate, hBN, or a polymer layer on the magneto-optical signals, \cite{Fang2018, Ma2019, Wu2019, Jin2020, Molina-Sanchez2020} a comprehensive study of thin-film interference effects for the magneto-optics in realistic samples is still lacking.

\begin{figure*}[htp]
    \centering
    \includegraphics[width=\linewidth]{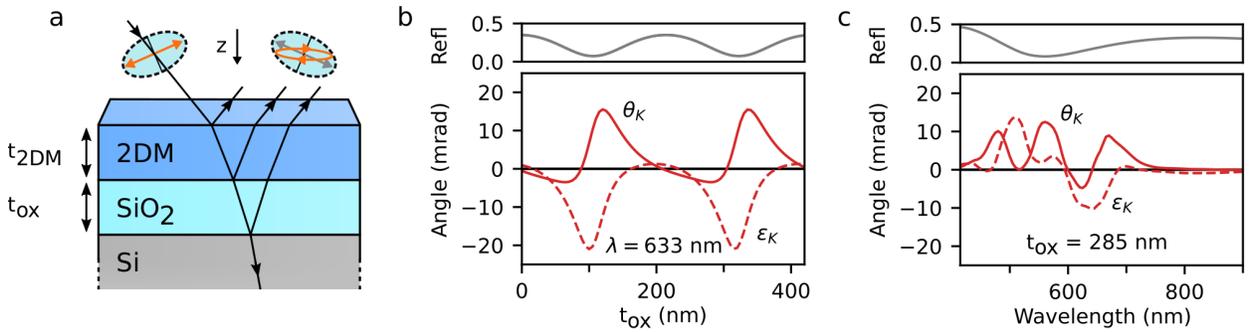}
    \caption{a) A typical 2D ferromagnet sample displaying MOKE in the presence of thin-film interference. b), c) The calculated Kerr rotation and ellipticity depend heavily on the oxide thickness and wavelength. The maximum signal occurs when the reflectivity is close to its minimum.}
    \label{fig:geometry_and_apetizers}
\end{figure*}

Here, we show that not only the substrate, but also other materials in a van der Waals stack can greatly affect the MOKE signals, and that these signals can be significantly enhanced by carefully choosing the illumination wavelength and through heterostructure engineering (Fig. \ref{fig:geometry_and_apetizers}), as is well known from other studies on thin-film interference enhancements of MOKE signals from e.g. metallic thin films.
Using a transfer matrix approach for thin-film interference, we demonstrate that the MOKE signals can reach values of tens to hundreds of milliradians at sizeable reflected light intensities.
In particular, we explore this effect on three systems based on the 2D van der Waals magnet CrI\ts{3} on a SiO\ts{2}/Si substrate: monolayer CrI\ts{3}, bulk CrI\ts{3}, and monolayer CrI\ts{3} encapsulated in hBN (we also consider other 2D magnets, see supplementary material).
Our results show that the often disregarded hBN encapsulation used to protect the air-sensitive 2D magnet films can strongly affect the magnitude of the MOKE signals, such that subtler magnetic textures in 2D magnets can be measured.

We model the thin layered systems as a series of stacked parallel homogeneous layers, where the first and last layer (being air and Si), are semi-infinite.
An example of this geometry is illustrated in Fig. \ref{fig:geometry_and_apetizers}a, where a single 2D magnetic layer of thickness $t_\text{2DM}$ is on top of a SiO\ts{2}/Si substrate with oxide thickness $t_\text{ox}$.
The interfaces are assumed to be smooth, such that there are only specular reflections.
Furthermore, we assume that the illumination intensity is low enough, such that the optical properties of the materials can be described by a linear dielectric permittivity tensor $\bm{\varepsilon}$ and magnetic permeability tensor $\bm{\mu}$.
The intensity and polarization of the light that reflects off this stratified linear system are calculated using the transfer matrix method. We use a method similar to the one in, \cite{Yeh1979} which is explained in full detail in the supplementary material.

The transfer matrix relates the components of the electric ($\vec{E}$) and magnetic ($\vec{H}$) field parallel to the layers, called $\vec{E}_\parallel$ and $\vec{H}_\parallel$, at one interface of a medium to the other one.
To construct a transfer matrix, we start by describing plane waves in a single layer.
We begin from the Maxwell equations in isotropic homogeneous media, and consider plane waves with a frequency $\omega$ and wave vector $\vec{k}$, of the form $\vec{E} = \vec{E}_0 \exp(i(\vec{k}\cdot \vec{r} - \omega t))$, where $t$ is time and $\vec{r}$ is the position in space.
We can then derive the following wave equation:
\begin{eqnarray}
\bm{\varepsilon}^{-1} \left(\vec{k} \times \left(\bm{\mu}^{-1} \left(\vec{k} \times \vec{E}_0\right)\right)\right) = -\omega^2\vec{E}_0. \label{eq:fresnel}
\end{eqnarray}
Solving the above equation yields four values for the $z$-component of $\vec{k}$, $k_{z,i}$, and four corresponding polarization eigenmodes, $\vec{E}_{0,i}$, where $i$ labels the polarization mode.
These solutions describe two plane waves traveling in the $+z$ direction, and two in the $-z$ direction.
The transfer matrix is the diagonal matrix $diag(\exp(ik_{z,i} t_\text{layer}))$, which propagates the eigenmodes with wave vector components $k_{z,i}$ from one interface to the other one over a distance $t_\text{layer}$, after it is transformed from the basis of the eigenmode amplitudes to the the basis of the amplitudes of the $\vec{E}_\parallel$ and $\vec{H}_\parallel$ components.
The transfer matrix of the whole system is simply the product of the transfer matrices of the individual layers, since $\vec{E}_\parallel$ and $\vec{H}_\parallel$ are continuous across the interfaces.
This matrix is used to calculate the amplitudes of the eigenmodes of the reflected and transmitted light, and from this the reflected intensity and polarization.

We apply the above method to the system illustrated in Fig. \ref{fig:geometry_and_apetizers}a, where the 2D ferromagnet is monolayer CrI\ts{3} with a thickness of $t_{2DM}$ = 0.7 nm.
The dielectric tensor of ferromagnetic monolayer CrI\ts{3} is taken from Wu et al.\cite{Wu2019}, where it is calculated from first-principles methods taking excitonic effects into account.
The dielectric constants of Si and thermally grown SiO\ts{2} are experimental values from Herzinger et al. \cite{Herzinger1998}
The magnetic permeability of all materials is approximated by the scalar vacuum permeability $\mu_0$.
Using these parameters, we calculate the Kerr angle $\theta_K$, Kerr ellipticity $\varepsilon_K$, and reflected intensity of linearly polarized light hitting the sample at normal incidence and polar configuration.
The results are shown in Fig. \ref{fig:geometry_and_apetizers}b and \ref{fig:geometry_and_apetizers}c as a function of $t_\text{ox}$ and wavelength respectively. 

Fig \ref{fig:geometry_and_apetizers}b shows a clear periodic behavior of the MOKE signals as a function of $t_\text{ox}$, with a period of 216 nm, corresponding to half a wavelength in SiO\ts{2}.
It also shows that the Kerr angle and ellipticity attain their maximum values when the reflected intensity is close to a minimum, and vice-versa.
In Fig. \ref{fig:geometry_and_apetizers}c, the largest MOKE signals are found in the wavelength range from 400 nm to 750 nm, where the wavelength dependence of $\theta_K$ and $\varepsilon_K$ is caused primarily by the wavelength dependence of the dielectric tensor of CrI$_3$.
Again, $\theta_K$ and $\varepsilon_K$ attain their maximum values when the reflectivity is close to a minimum.
These results show that the oxide thickness and the wavelength of the light have a strong impact on the sign and magnitude of the MOKE signals.
By optimizing $t_\text{ox}$ or the wavelength, the signals can already change by as much as 20 mrad in this example, while still having a sizable reflectivity of more than 6\%.

\begin{figure}
    \centering
    \includegraphics[]{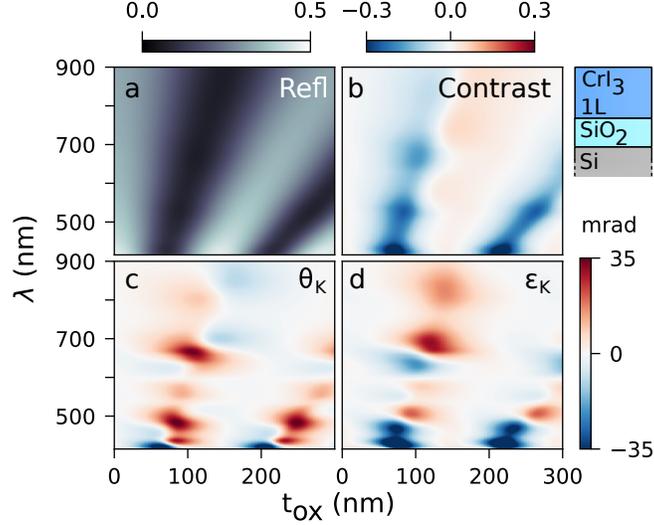}
    \caption{Simulation results for a CrI\ts{3}(1L)-SiO\ts{2}-Si stack. The reflectivity (a), contrast (b), Kerr rotation (c) and ellipticity (d) are shown as function of illumination wavelength and oxide thickness. Where the color scale is saturated, the values exceed the bounds of the scale.}
    \label{fig:SiO2_thickness_influence}
\end{figure}

In order to get a complete picture of the impact of each parameter on the signals, we explore the full parameter space, varying both the wavelength and oxide thickness for a CrI\ts{3} monolayer on a SiO\ts{2}/Si substrate (Fig. \ref{fig:SiO2_thickness_influence}).
Besides the reflectivity, $\theta_K$, and $\varepsilon_K$, we also calculate the contrast for the CrI\ts{3} layer. This can be used to locate the target flake, usually a few $\mu$m in size and therefore hard to find on a large substrate, using a microscope, or using a reflectivity scan in a laser based experiment.
The contrast is defined as $C = (I - I_0) / I_0$, where $I$ and $I_0$ are the reflected intensity of the system with and without CrI\ts{3} respectively.
The reflectivity in Fig. \ref{fig:SiO2_thickness_influence} shows a clear fan pattern.
The periodicity in $t_\text{ox}$ in our simulated reflectivity corresponds to half a wavelength in the SiO\ts{2}, which strongly suggests that this fan pattern is caused by the interference of the light reflected from the top and bottom interface of the SiO\ts{2}, similar to graphene-based systems. \cite{Blake2007}
The same pattern appears for $C$, $\theta_K$, and $\varepsilon_K$, indicating that the interference in the SiO\ts{2} layer also has a large effect on the contrast and MOKE signals.
Additional features at 420 nm, 500 nm, and 680 nm, can also be seen, and originate from the wavelength dependence of the dielectric tensor of CrI\ts{3} (see supplementary material).
By tuning both the wavelength and oxide thickness, $\theta_K$ and $\varepsilon_K$ can be tuned over a range of several tens of milliradians while keeping the reflectivity above 5\%.
Furthermore, when the Kerr rotation and ellipticity are maximized, the contrast is large as well, making it easier to locate the CrI\ts{3} using e.g. a simple reflectivity scan.

The above results can be compared to the experimental results from Huang et al. \cite{Huang2017}
In their experimental work, using a laser with a wavelength of 633 nm and $t_\text{ox}$ = 285 nm, they obtained $\theta_K = 5 \pm 2$ mrad.
Our theoretical result of 3.5 mrad is within the experimental error margin.
Our results are also in agreement with the absence of an experimental signal at a wavelength of 780 nm for this system.
We find that the MOKE signals at these wavelengths are reduced by about a factor of ten and could easily be obscured by the experimental noise.
Fig. \ref{fig:geometry_and_apetizers} also indicates that the combination of an oxide thickness of 285 nm and a laser wavelength of 633 nm does not result in the largest Kerr rotation.
Using instead an oxide thickness of 335 nm would increase $\theta_K$ by more than a factor of 4, or if the wavelength is changed to 560 nm, the Kerr rotation can increase by a factor of about 3.

\begin{figure}
    \centering
    \includegraphics[]{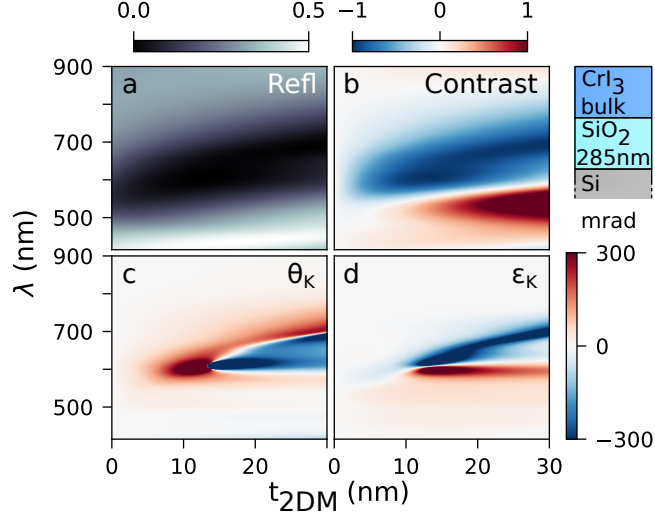}
    \caption{Simulation results for a CrI\ts{3}(bulk)-SiO\ts{2}(285 nm)-Si stack. The reflectivity (a), contrast (b), Kerr rotation (c) and ellipticity (d) are shown as function of wavelength and CrI\ts{3} thickness. Where the color scale is saturated, the values exceed the bounds of the scale.}
    \label{fig:CrI3_thickness_influence}
\end{figure}

The 2DM thickness can also strongly affect the MOKE signals.
Fig. \ref{fig:CrI3_thickness_influence} shows the dependence of the magneto-optical signals as a function of both wavelength and 2DM thickness, using the dielectric tensor of ferromagnetic bulk CrI\ts{3} taken from Wu et al.\cite{Wu2019}
While the theoretical values of $\varepsilon_{\text{CrI}_3}$ used in our calculations differ slightly from the available experimental values, \cite{Grant1968, Huang2017} our main findings are not altered if we consider the experimental values.
We therefore opt for using the theoretical values since they span a larger wavelength range.
For comparison, we provide calculations using the experimental values in the supplementary material.
Interestingly, $\theta_K$ and $\varepsilon_K$ have a non-monotonic behavior, showing a strong peak and dip around a wavelength of 600 nm CrI\ts{3} thickness of 14 nm.
The extreme values of $\theta_K$ and $\varepsilon_K$ approach $\pm \pi/2$ and $\pm \pi/4$ respectively. The reflectivity at these points is around 0.2\% for the maximum Kerr rotation and 0.4\% for the maximum Kerr ellipticity.
These extreme MOKE signals can therefore be very hard to detect.
However, $\theta_K$ and $\varepsilon_K$ can still be changed over a range of a few hundred milliradians when tuning the wavelength and CrI\ts{3} thickness, while keeping the reflectivity above 5\% and having a good contrast.

\begin{figure}
    \centering
    \includegraphics[]{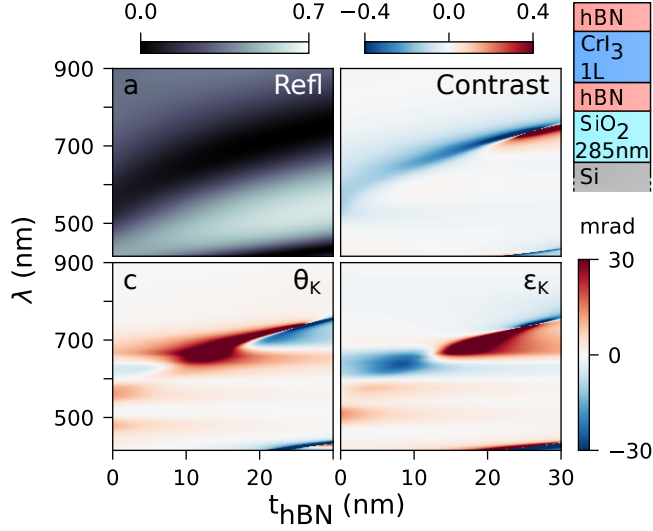}
    \caption{Simulation results for a hBN-CrI\ts{3}(1L)-hBN-SiO\ts{2}(285 nm)-Si stack. The reflectivity (a), contrast (b), Kerr rotation (c) and ellipticity (d) are shown as function of wavelength and hBN thickness. Where the color scale is saturated, the values exceed the bounds of the scale.}
    \label{fig:hbn_thickness_influence}
\end{figure}

Due to the air sensitivity of many 2DMs, they are often encapsulated in hBN. \cite{Huang2018, Wang2018a}
The presence of the hBN layers also leads to thin-film interference effects and thus can be used to engineer the magneto-optical signals as well. \cite{Jin2020}
To explore the impact of hBN encapsulation, we study the MOKE signals in monolayer CrI\ts{3} encapsulated by a top and bottom hBN flake with the same thickness $t_\text{hBN}$.
The refractive index of hBN needed for the simulation is calculated using the single oscillator model, $n(\lambda)
^2= 1 + A\lambda^2 / (\lambda^2 - \lambda_0^2)$, where $\lambda_0$ = 164.4 nm and $A$ = 3.263 are determined experimentally by Lee et al. \cite{Lee2019}
The simulation results for an oxide layer of 285 nm are shown in Fig. \ref{fig:hbn_thickness_influence}.
We have also investigated the effect of the hBN thickness on the signal-to-noise ratio (see supplementary material).
A striking result is that an hBN thickness of about ten nanometers, a typical thickness for hBN flakes used for encapsulation in experimental studies, can already lead to dramatic changes in the reflectivity, contrast, and Kerr signals.
Therefore, one should take into account the system as a whole when engineering their heterostructures for optimal MOKE signals.
The hBN encapsulation is particularly important, since the wavelength and oxide thickness are usually more difficult to vary, while hBN flakes of various thicknesses can be easily found in a single exfoliation run.
Therefore, in addition to protecting the 2DM against degradation, hBN encapsulation can be used as an active method for magneto-optical signal enhancement.

The common feature in the results of the simulation of the three systems above that $\theta_K$ and $\varepsilon_K$ are maximized when the reflectivity is close to a minimum, is a general and well-known phenomenon. \cite{Mansuripur1982, Zak1990, Bass2010} It can be explained by the behavior of the reflection coefficients for the electric field of the two circular polarizations,  $r_+$ and $r_-$, near the reflectivity minimum.
In this region, the magnitude of both reflection coefficients are small, and their complex phases change rapidly with wavelength and layer thickness.
The exact parameter values around which these coefficients have a minimum and change phase are different for $r_+$ and $r_-$ due to the circular birefingence and dichroism caused by the magnetic layer.
Therefore, both the ellipticity, given by $\varepsilon_K = \tan^{-1}(|r_+| - |r_-|) / (|r_+| + |r_-|)$, and the Kerr rotation, given by $\theta_K = \left(arg(r_+) - arg(r_-)\right) / 2$, can become very large when the total reflectivity is near a minimum, as is explained in more detail in the supplementary material.
On the other hand, if the reflectivity is large, both $r_+$ and $r_-$ are large, meaning that their relative difference is small, and that their complex phase changes slowly with wavelength and layer thickness. This will result in a low Kerr ellipticity and rotation respectively. Therefore, the extreme MOKE signals of e.g. $\varepsilon_K \approx \pm \pi / 4$ and $\theta_K \approx \pi / 2 $ calculated for the CrI$_3$(bulk)-SiO$_2$(285nm)-Si stack can only occur at a low reflectivity.
This reasoning is not restricted to the samples treated in this paper.
A general method to increase the Kerr rotation and ellipticity of a multi-layer sample is to use a combination of wavelength and thickness of the layers that minimizes the reflectivity.
A reduction of the reflectivity, and a corresponding increase the magneto-optical signals, can also be achieved by adding new layers to the sample.
Such anti-reflection coatings have been used for over half a century to enhance Kerr signals from magnetic films. \cite{Kranz1958, Mansuripur1982, Cantwell2006, Kim2020}

Here we showed that thin-film interference can be a useful tool for improving magneto-optical signals in magnetic van der Waals systems.
Through careful sample or heterostructure engineering, one is able to optimize their system for a particular experimental setup, improving the signal-to-noise ratio and measurement speed.
The optimization of the signals can be done by choosing a particular illumination wavelength, substrate, thickness of the van der Waals magnet, or hBN used for encapsulation.
The signal improvement, reaching several tens of miliradians, could lead to the identification of weaker signals from more delicate effects, such as chiral magnetic structures embedded in a homogeneously magnetized lattice.


\section*{Supplementary Material}
See supplementary material for the simulation details, graphs of the dielectric tensors used in the simulations, simulation results for other 2D magnetic monolayers, and an explanation for why the Kerr rotation and ellipticity are large when the reflectivity is close to a minimum.

\begin{acknowledgments}
We thank Alejandro Molina-S\'anchez, for sharing their data on the dielectric tensor for chromium trihalides.
This work was supported by the Zernike Institute for Advanced Materials, the Dutch Research Council (NWO Start-Up, STU.019.014), and the European Union’s Horizon 2020 research and innovation programme under grant agreement No 785219 (Graphene Flagship Core 3).
\end{acknowledgments}

\section*{Data availability}
The data that support the findings of this study are available from the corresponding author upon reasonable request.

\bibliographystyle{aapmrev4-1}
\bibliography{manuscript.bib}

\end{document}


\title{Supplementary material of 'Enhancing magneto-optic effects in two-dimensional magnets by thin-film interference'} 

\author{F. Hendriks}
\email[]{f.hendriks@rug.nl}
\affiliation{Zernike Institute for Advanced Materials, University of Groningen, The Netherlands}
\author{M.H.D. Guimarães}
\email[]{m.h.guimaraes@rug.nl}
\affiliation{Zernike Institute for Advanced Materials, University of Groningen, The Netherlands}

\date{\today}

\maketitle 

\tableofcontents
\newpage

\section{Simulation details}
\label{appendix:simulation_details}
The thin layered systems we simulate in this work are modelled by a series of stacked parallel homogeneous layers, where the first and last layers (being air and silicon) are semi-infinite.
An example of this geometry is depicted in Fig. 1a of the main text.
The interfaces are assumed to be smooth, such that all reflections are perfectly specular.
Furthermore, the illuminating intensity is assumed to be low enough, such that the optical properties of the materials are described by a linear dielectric permittivity tensor ($\bm{\varepsilon}$) and a linear magnetic permeability tensor ($\bm{\mu}$).
To calculate the intensity and polarization of the light that reflects off this stratified medium, we use a slight variation of the transfer matrix method described by P. Yeh et al. \cite{Yeh1979}, where we assume a more general wave equation and use a different basis for the transfer matrix.
This method is explained in detail below.

A transfer matrix relates the components of the electric ($\vec{E}$) and magnetic ($\vec{H}$) field parallel to the layers, $\vec{E}_\parallel$ and $\vec{H}_\parallel$ respectively, of the bottom and top interfaces of a medium.
Grouping these parallel field components into a vector $\vec{F}$, the transfer matrix $M$ is defined as
\begin{eqnarray}
\vec{F}^{II} = M \vec{F}^{I}, \quad \vec{F}^{i} =
\begin{bmatrix}
E^{i}_{x} \\
E^{i}_{y} \\
H^{i}_{x} \\
H^{i}_{y}
\end{bmatrix},
\label{eq:F_def}
\end{eqnarray}
where the superscript $i$ = $I$, $II$, specifies the interface.
To construct a transfer matrix, we start by assuming plane waves propagating in a layer:
\begin{align}
    \vec{E}(\vec{r}, t) &= \vec{E_0}e^{i\left(\vec{k}\cdot\vec{r} - \omega t\right)},
    \label{eq:planewave}
\end{align}
where $\vec{E}_0$ is the electric field amplitude, $\vec{k}$ the wave vector, $\vec{r}$ the position in space, $\omega$ the angular frequency, and $t$ time.
Plugging this into Maxwell's equations gives
\begin{subequations}
    \begin{eqnarray}
    \vec{k} \cdot \left(\bm{\varepsilon} \vec{E}\right) &= 0 \label{eq:gauss}\\
    \vec{k} \times \vec{E} &= \omega \vec{B}_0 \label{eq:faraday}\\
    \vec{k} \cdot \vec{B} &= 0 \label{eq:div_B}\\
    \vec{k} \times \left( \bm{\mu}^{-1}\vec{B}\right) &= -\omega\bm{\varepsilon}\vec{E}. \label{eq:ampere}
    \end{eqnarray}
\end{subequations}
Using Eq. \ref{eq:faraday} to eliminate $\vec{B}$ from Eq. \ref{eq:ampere} yields
\begin{eqnarray}
\bm{\varepsilon}^{-1} \left(\vec{k} \times \left(\bm{\mu}^{-1} \left(\vec{k} \times \vec{E}_0\right)\right)\right) = -\omega^2\vec{E}_0. \label{eq:fresnel_A}
\end{eqnarray}
This equation gets a more familiar form when the cross products are written in terms of matrix multiplications. Defining
\begin{align}
    \bm{k}_{c.p.} = 
    \begin{bmatrix}
    0 & -k_z & k_y\\
    k_z & 0 & -k_x\\
    -k_y & k_x & 0
    \end{bmatrix},
\end{align}
Eq. \ref{eq:fresnel_A} becomes
\begin{align}
    \bm{\varepsilon}^{-1} \bm{k}_{c.p.} \bm{\mu}^{-1} \bm{k}_{c.p.} \vec{E}_0 = -\omega^2\vec{E}_0. \label{eq:fresnel_A2}
\end{align}
Now it takes the form of an eigenvalue problem for a matrix given by $\bm{\varepsilon}^{-1} \bm{k}_{c.p.} \bm{\mu}^{-1} \bm{k}_{c.p.}$ with eigenvalue $-\omega^2$.
Since $E_\parallel$ and $H_\parallel$ are continuous across all interfaces, $\omega$, $k_x$ and $k_y$ are constant throughout the system, and therefore they are equal to $\omega$, $k_x$ and $k_y$ of the incoming light.
The only unknowns are $k_z$ and $\vec{E}_0$. 
The solution for this eigenvalue problem gives a relation between $\vec{k}$, $\omega$, and the material parameters $\bm{\varepsilon}$ and $\bm{\mu}$.
This yields four values for the $z$-component of $\vec{k}$, $k_{z,i}$, describing two waves traveling in the $+z$ and two in the $-z$ direction.
The corresponding eigenvectors, $E_{0,i}$, are the polarization modes of the electric field.
For the magnetic field, the polarization modes are $H_{0,i} = \bm{\mu}^{-1}(\vec{k} \times \vec{E}_0) / \omega$.
This describes plane wave propagation in a single linear homogeneous medium.

The transfer matrix for a single layer can be constructed from the polarization modes and the wave vectors of the four plane waves that are allowed for a given incident wave.
Given $E_\parallel$ and $H_\parallel$ at one interface, their value at the other interface of the medium is calculated by decomposing the fields in the polarization modes, propagating these modes to the other interface using Eq. \ref{eq:planewave}, and then transforming them back to $E_\parallel$ and $H_\parallel$.
The transfer matrix relating $E_\parallel$ and $H_\parallel$ between the two interfaces is
\begin{align}
M = &A P A^{-1},
\end{align}
where the matrices $A$ and $P$ are defined as
\begin{align}
A = 
&\begin{bmatrix}
E_{0,x1} & E_{0,x2} & E_{0,x3} & E_{0,x4}\\
E_{0,y1} & E_{0,y2} & E_{0,y3} & E_{0,y4}\\
H_{0,x1} & H_{0,x2} & H_{0,x3} & H_{0,x4}\\
H_{0,y1} & H_{0,y2} & H_{0,y3} & H_{0,y4}\\
\end{bmatrix}, \quad
P = 
\begin{bmatrix}
e^{i\left(k_{z,1}\Delta z\right)} & 0 & 0 & 0\\
0 & e^{i\left(k_{z,2}\Delta z\right)} & 0 & 0\\
0 & 0 & e^{i\left(k_{z,3}\Delta z\right)} & 0\\
0 & 0 & 0 & e^{i\left(k_{z,4}\Delta z\right)}\\
\end{bmatrix}.
\end{align}
Since $E_\parallel$ and $H_\parallel$ are continuous across the interfaces, the transfer matrix for light propagation through whole system is described by the product of the transfer matrices of the individual layers:
\begin{align}
    M_{\text{tot}} = \prod_{j}M_{j}.
\end{align}
This matrix relates $E_\parallel$ and $H_\parallel$ in the first layer to $E_\parallel$ and $H_\parallel$ in the final layer.

We now write $E_\parallel$ and $H_\parallel$ as linear combinations of the polarization modes ($\vec{F}_i$), with a coefficient ($a_i$):
\begin{align}
    \vec{F} = \sum_i a_i \vec{F}_i.
\end{align}
For the final layer, we assume that there are only waves traveling in the $+z$ direction, since there is no interface to reflect them back.
Substituting the above relation in Eq. \ref{eq:F_def} and rearranging the terms results in:
\begin{align}
a_{i1}M_{\text{tot}}\vec{F}_{i1} + a_{i2}M_{\text{tot}}\vec{F}_{i2} + a_{r1}M_{\text{tot}}\vec{F}_{r1} + a_{r2}M_{\text{tot}}\vec{F}_{r2} =
a_{t1}\vec{F}_{t1} + a_{t2}\vec{F}_{t2},
\end{align}
where the subscripts $i$, $r$, and $t$ label the incident, reflected, and transmitted modes respectively.
For example, $\vec{F}_{r1}$ and $\vec{F}_{r2}$ represent the first and second polarization mode of the reflected beam ($-k_z$). 

After regrouping the terms, the (complex) amplitudes of the reflected and transmitted modes can be solved from the following matrix equation:
\begin{align}
    \begin{bmatrix}
    M_\text{tot}\vec{F}_{r1} & M_\text{tot}\vec{F}_{r2} & \vec{F}_{t1} & \vec{F}_{t2}
    \end{bmatrix}
    \begin{bmatrix}
    a_{r1}\\
    a_{r2}\\
    a_{t1}\\
    a_{t2}
    \end{bmatrix}
    =
    M_\text{tot}
    \begin{bmatrix}
    \vec{F}_{i1} & \vec{F}_{i2}
    \end{bmatrix}
    \begin{bmatrix}
    a_{i1}\\
    a_{i2}
    \end{bmatrix}.
\end{align}

These complex amplitudes describe the intensity and the polarization of the reflected and transmitted waves.
Since they travel in an isotropic medium, the eigenmodes can be chosen to be the $p$ and $s$ polarization modes, or horizontal and vertical polarization modes in case the incoming wave is perpendicular to the layers.
The Jones vector is then calculated as $J = [a_1, a_2] / (|a_1|^2 + |a_2|^2)$.
Considering only the phase difference ($\Delta$) between $a_1$ and $a_2$ and writing the Jones vector in the form $[a, b \exp(i\Delta)]$, where $a$ and $b$ are positive real numbers, the polarization angle $\theta$ and ellipticity $\varepsilon$ are calculated by:
\begin{align}
    \tan(\theta) &= 2ab\cos(\Delta) / (a^2 - b^2)\\
    \tan(\varepsilon) &= \sqrt{(1-q)/(1+q)}, \quad
    q = \sqrt{1-(2ab\sin{\Delta})^2}.
\end{align}
We consider the ellipticity being positive when $\sin(\Delta) < 0$.
The difference between the $\theta$ and $\varepsilon$ of the reflected (transmitted) and initial wave gives the Kerr (Faraday) rotation and ellipticity.

\section{Dielectric tensors of the materials used in the simulations}
This section shows the values of the dielectric tensor elements of the materials used in our simulations.
The dielectric tensor elements are plotted versus wavelength for the range we consider in the main text, which is from 414 nm to 900 nm.
We refer to the respective original publications (referenced in each figure caption) for the full data.

\begin{figure*}[htp]
    \centering
    \includegraphics[width=0.8\textwidth]{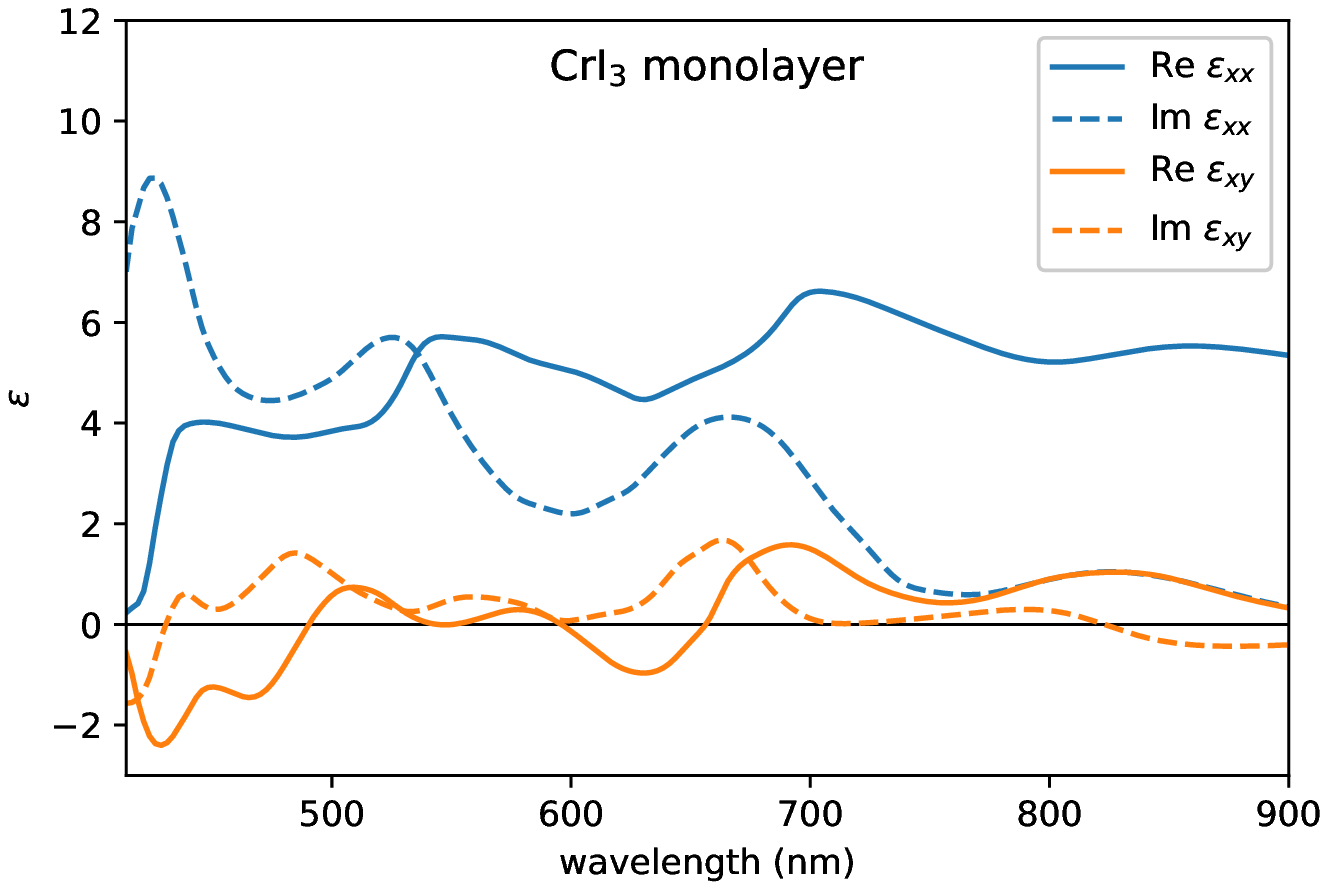}
    \caption{Dielectric tensor elements of monolayer CrI\ts{3}, calculated by Wu et al. \cite{Wu2019}}
    \label{fig:supp_eps_CrI3_ml_Wu2019}
\end{figure*}

\begin{figure*}[htp]
    \centering
    \includegraphics[width=0.8\textwidth]{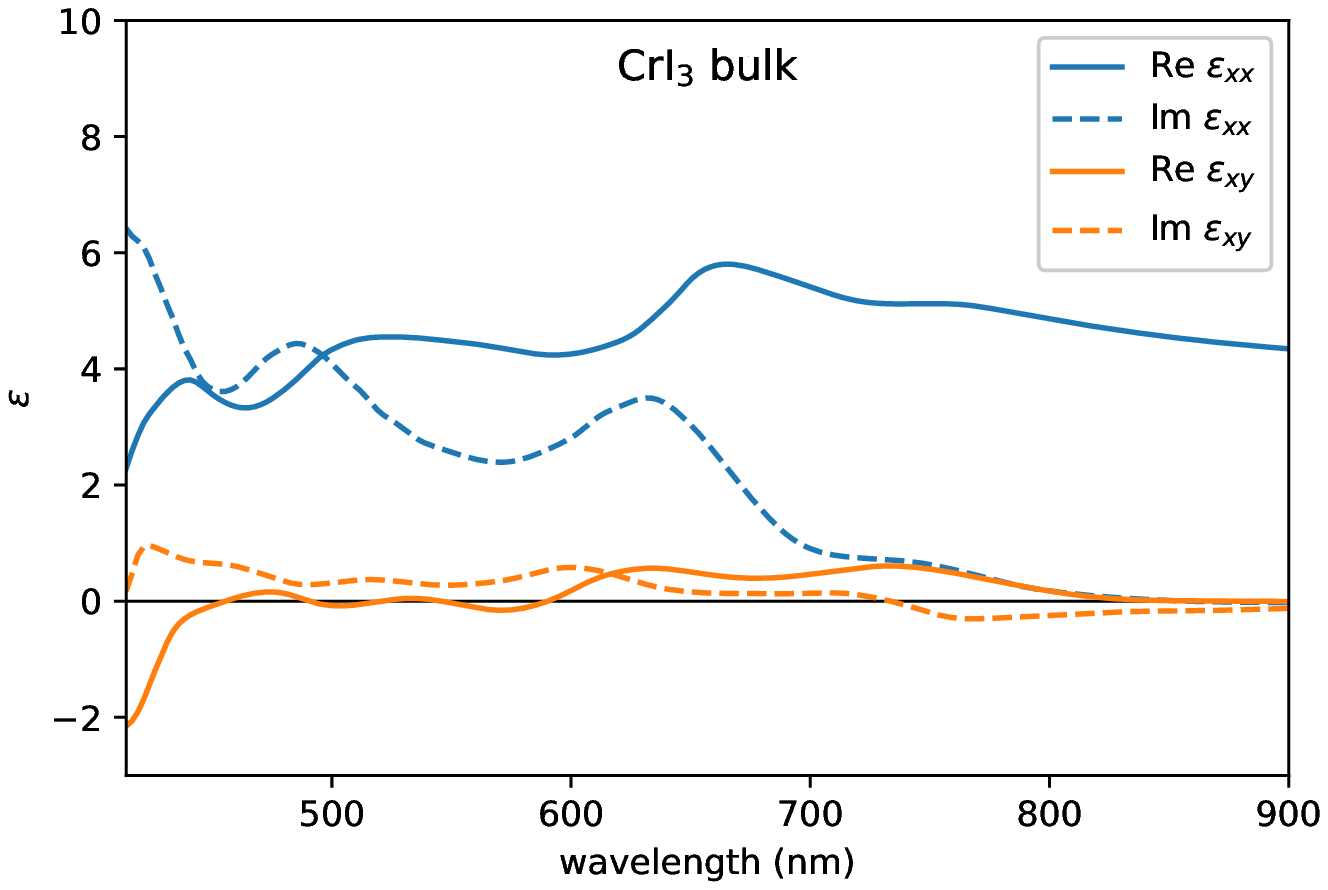}
    \caption{Dielectric tensor elements of bulk CrI\ts{3}, calculated by Wu et al. \cite{Wu2019}}
    \label{fig:supp_eps_CrI3_bulk_Wu2019}
\end{figure*}

\begin{figure*}[htp]
    \centering
    \includegraphics[width=0.8\textwidth]{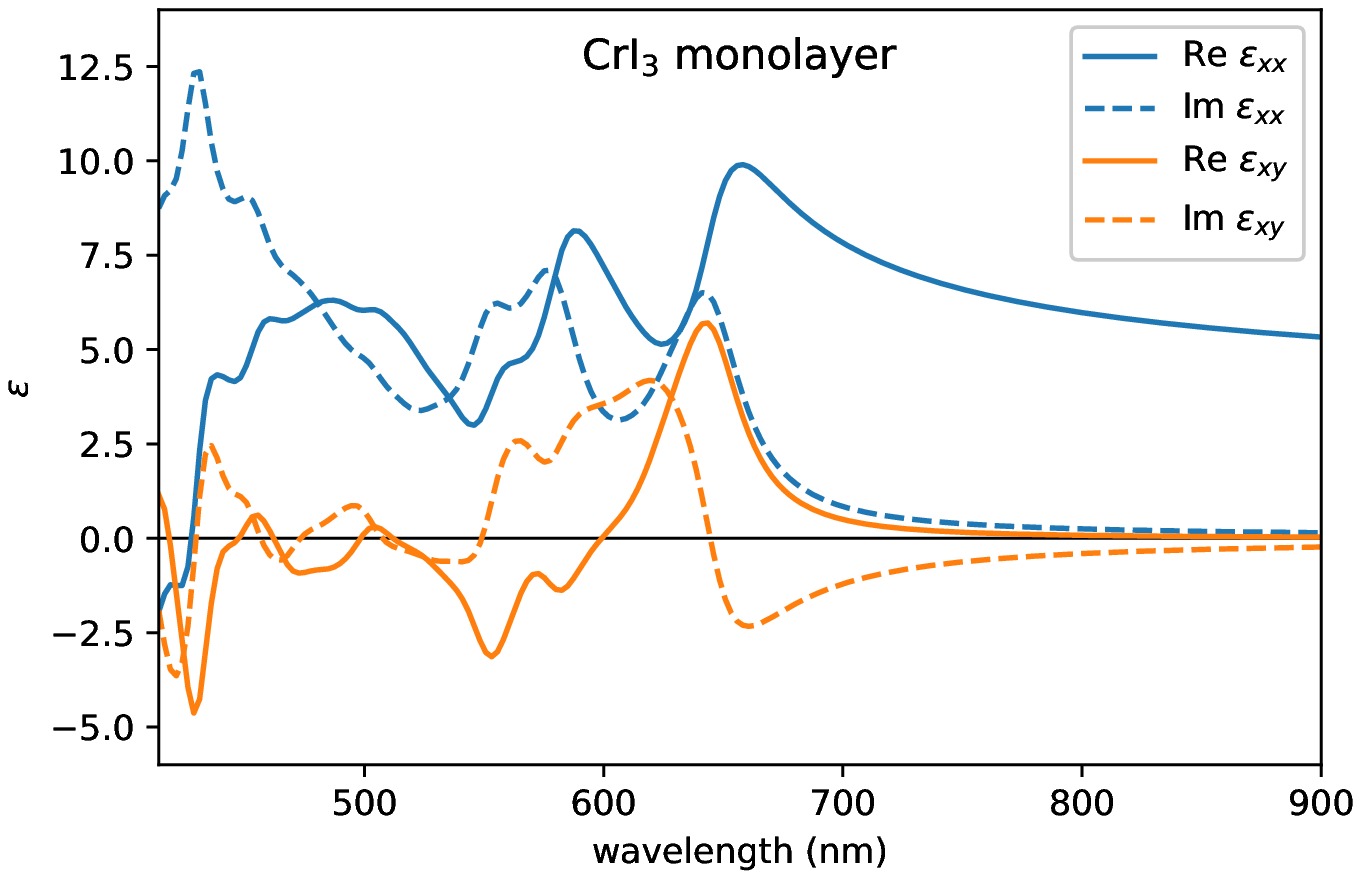}
    \caption{Dielectric tensor elements of monolayer CrI\ts{3}, calculated by Molina-Sánchez et al. \cite{Molina-Sanchez2020}}
    \label{fig:supp_eps_CrI3_ml_MolinaSanchez2020}
\end{figure*}

\begin{figure*}[htp]
    \centering
    \includegraphics[width=0.8\textwidth]{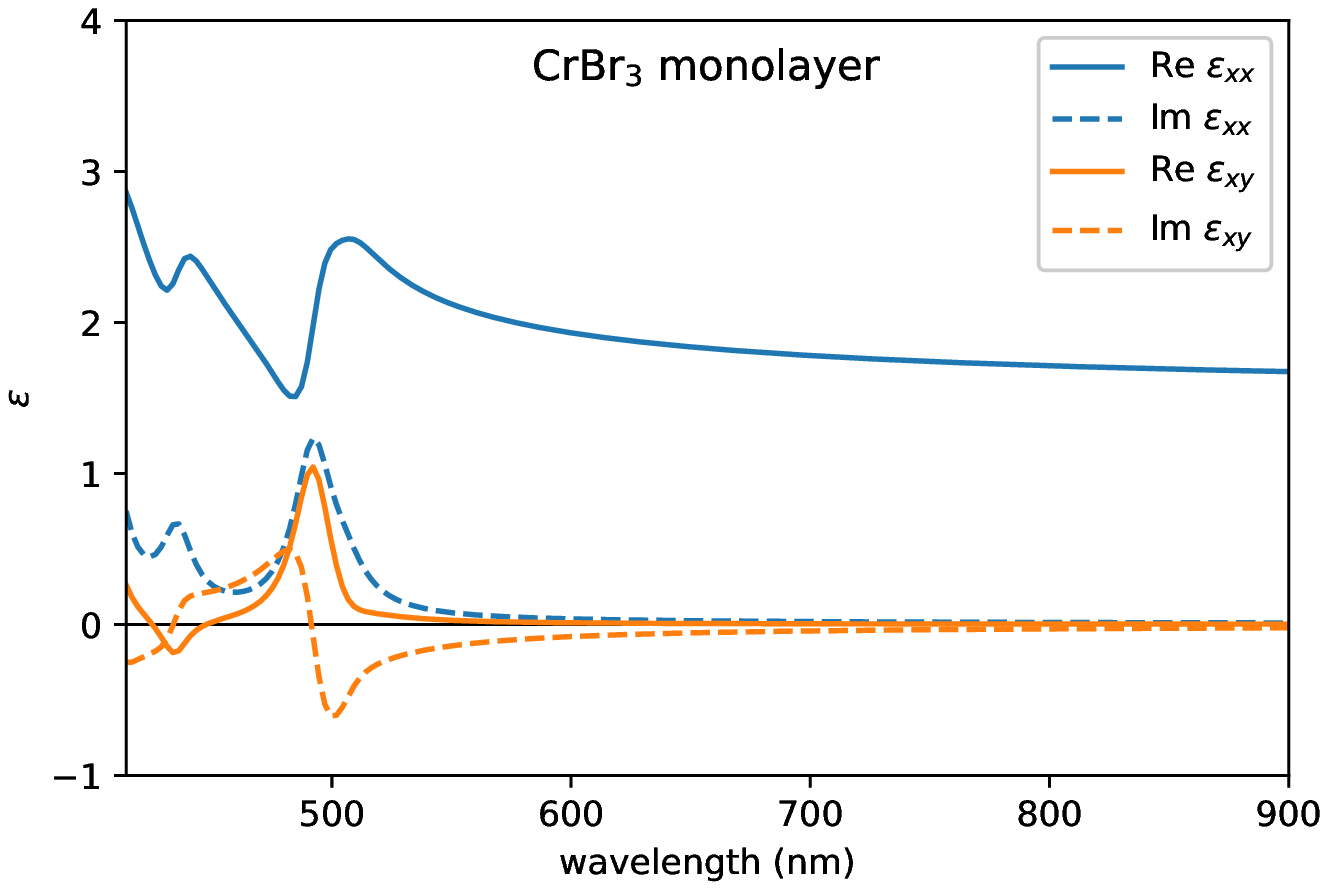}
    \caption{Dielectric tensor elements of monolayer CrBr\ts{3}, calculated by Molina-Sánchez et al. \cite{Molina-Sanchez2020}}
    \label{fig:supp_eps_CrBr3_ml_MolinaSanchez2020}
\end{figure*}

\begin{figure*}[htp]
    \centering
    \includegraphics[width=0.8\textwidth]{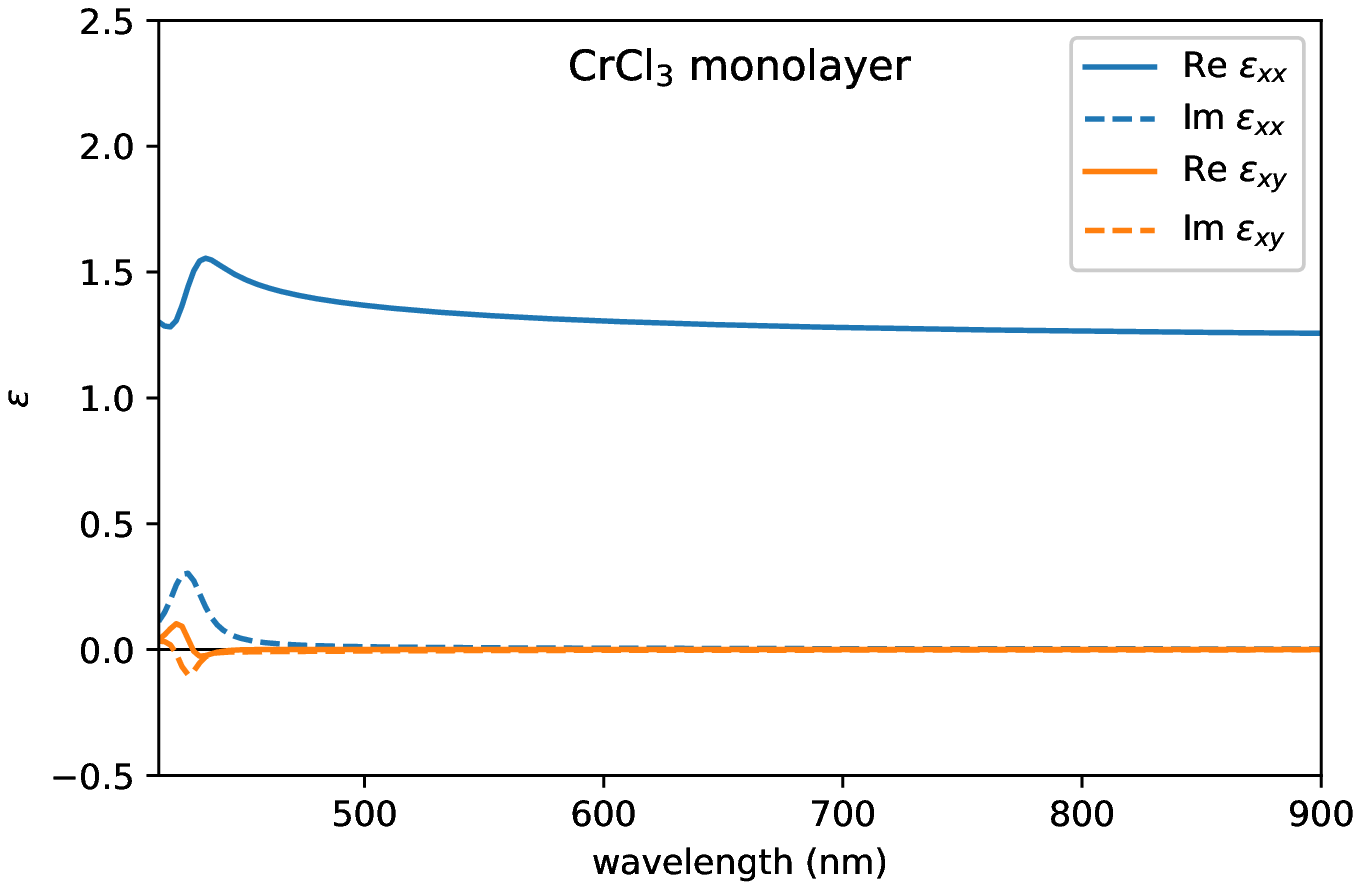}
    \caption{Dielectric tensor elements of monolayer CrCl\ts{3}, calculated by Molina-Sánchez et al. \cite{Molina-Sanchez2020}}
    \label{fig:supp_eps_CrCl3_ml_MolinaSanchez2020}
\end{figure*}

\begin{figure*}[htp]
    \centering
    \includegraphics[width=0.8\textwidth]{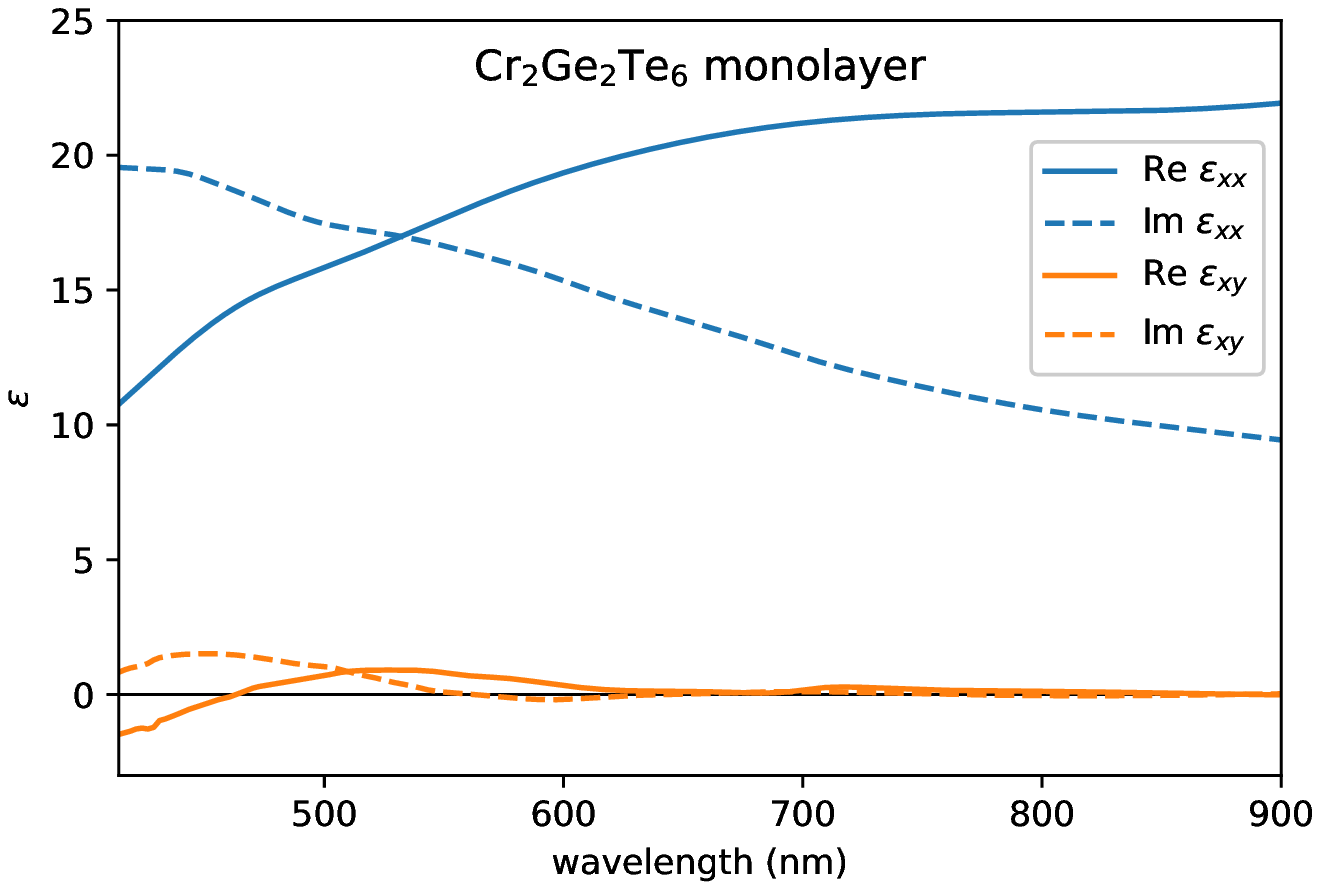}
    \caption{Dielectric tensor elements of monolayer C\ts{2}Ge\ts{2}Te\ts{6}, obtained from the optical conductivity calculated by Fang et al. \cite{Fang2018}}
    \label{fig:supp_eps_Cr2Ge2Te6_ml_Fang2018}
\end{figure*}

\begin{figure*}[htp]
    \centering
    \includegraphics[width=0.8\textwidth]{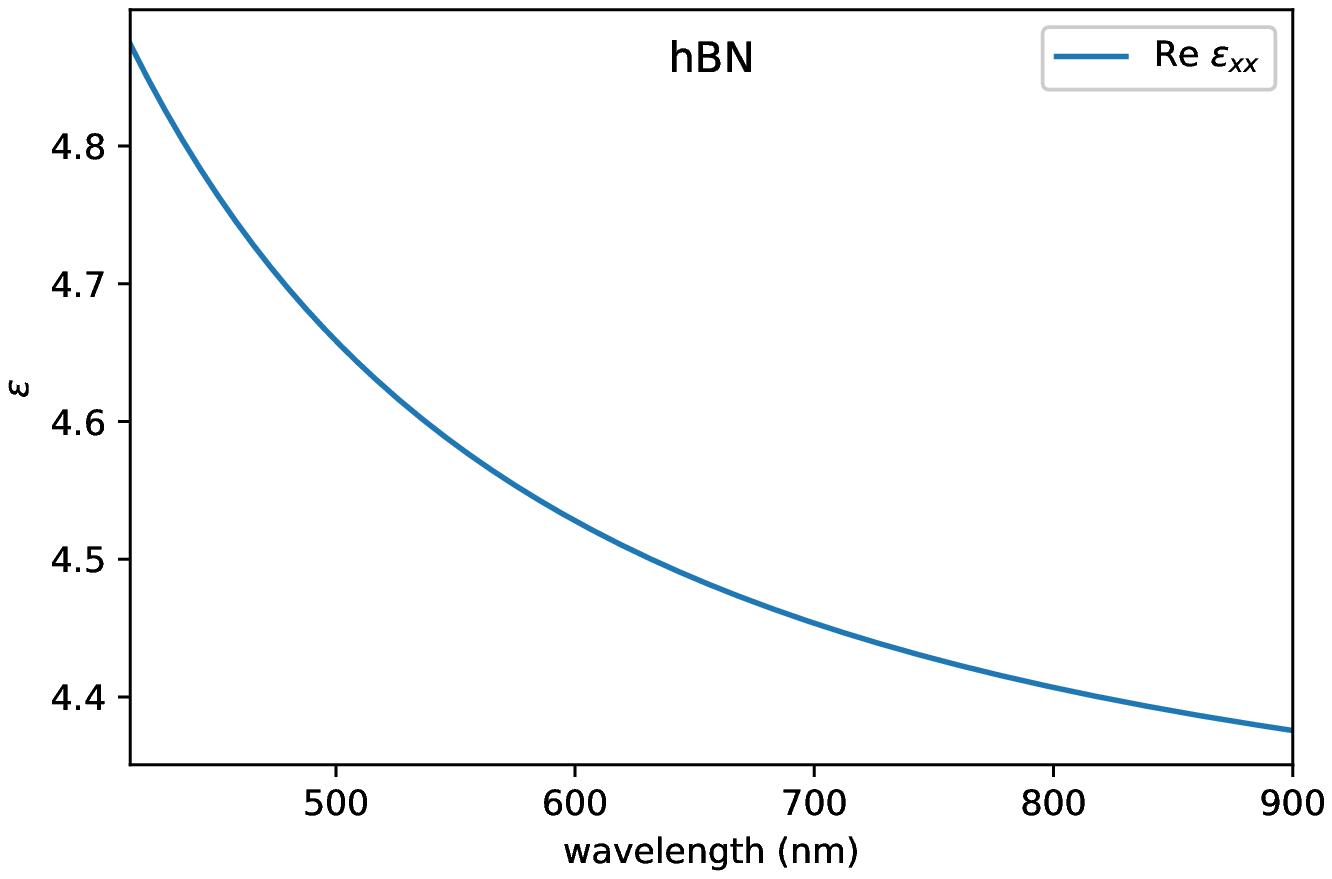}
    \caption{Dielectric constant of bulk hexagonal boron nitride (hBN), obtained from the refractive index measured by Lee et al. \cite{Lee2019} The refractive index is described by the single oscillator model $n = 1 + A\lambda^2 / (\lambda^2 - \lambda_0^2)$, with parameter values $A = 3.263$ and $\lambda_0 = 164.4$ nm.}
    \label{fig:supp_eps_hBN_Lee2019}
\end{figure*}

\begin{figure*}[htp]
    \centering
    \includegraphics[width=0.8\textwidth]{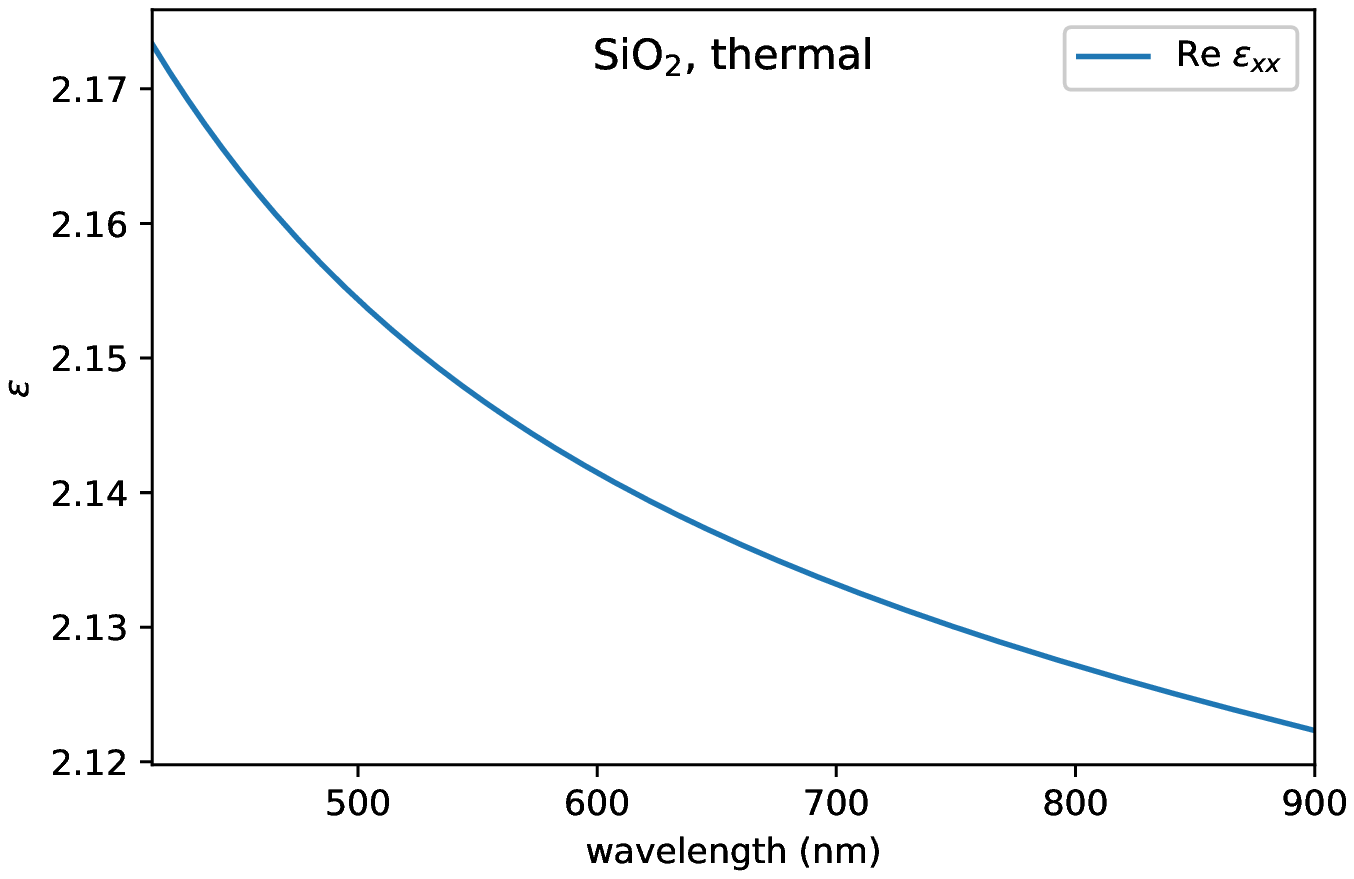}
    \caption{Dielectric constant of thermally grown SiO\ts{2}, obtained from the refractive index measured by Herzinger et al. \cite{Herzinger1998} It is described by the function $\varepsilon = n^2 = \text{offset} + a \lambda^2 / (\lambda^2 - b^2) - c\lambda^2$, with paramteter values $\text{offset} = 1.3000$, $a = 0.81996$, $b=0.10396$ $\mu\text{m}$, and $c=0.01082$.}
    \label{fig:supp_eps_SiO2_Herzinger1998}
\end{figure*}

\begin{figure*}[htp]
    \centering
    \includegraphics[width=0.8\textwidth]{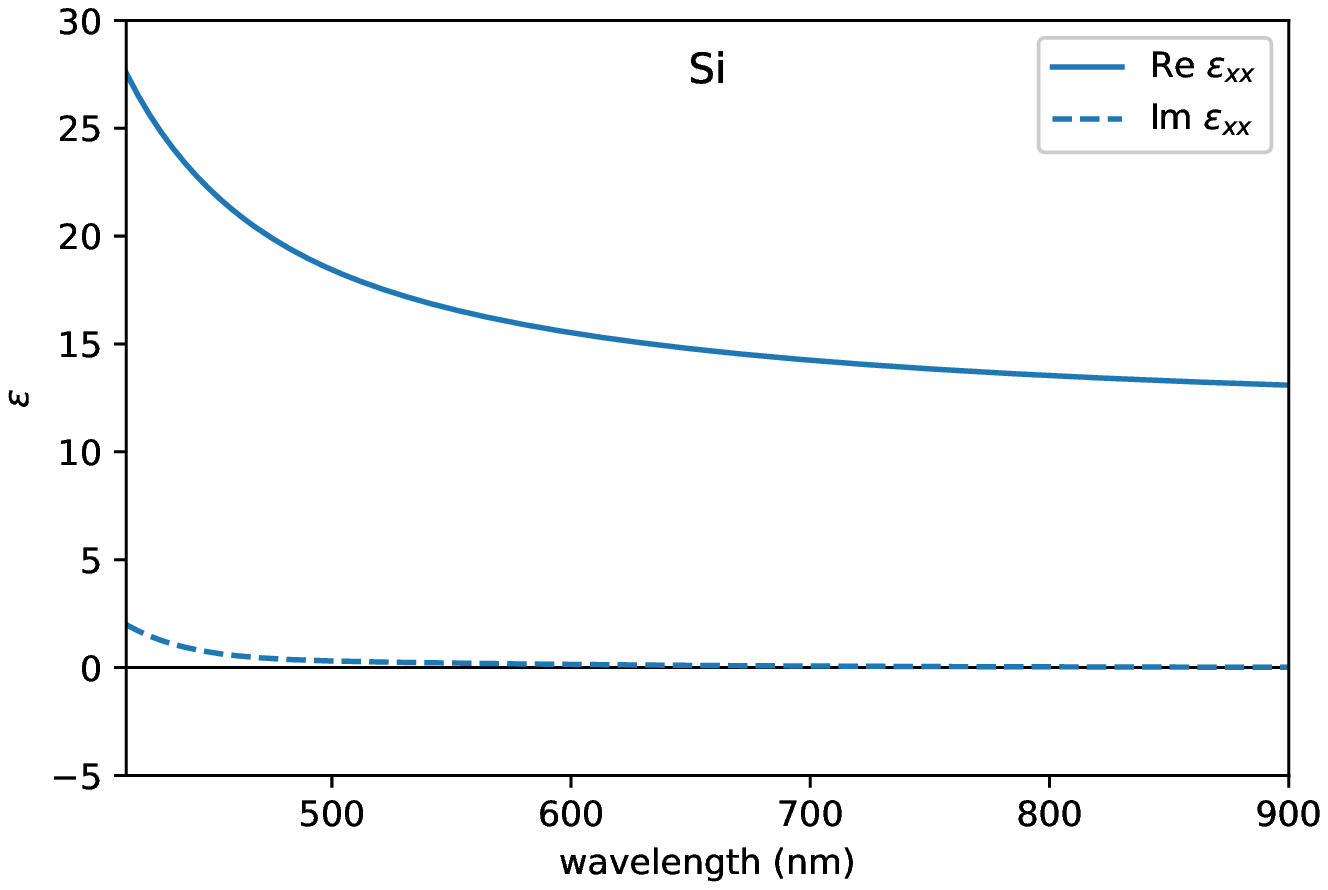}
    \caption{Dielectric constant of Si, measured by Herzinger et al. \cite{Herzinger1998}}
    \label{fig:eps_Si_Herzinger1998}
\end{figure*}

\FloatBarrier
\section{Interference effects in other 2D magnets}
\label{S_sec:other_magnets}
In this section we present the simulation results for heterostructures composed of different 2D magnetic monolayers, with a similar configuration as the one shown in Fig. 2 of the main text for monolayer CrI$_3$.

\begin{figure*}[htp]
    \centering
    \includegraphics[width=0.75\textwidth]{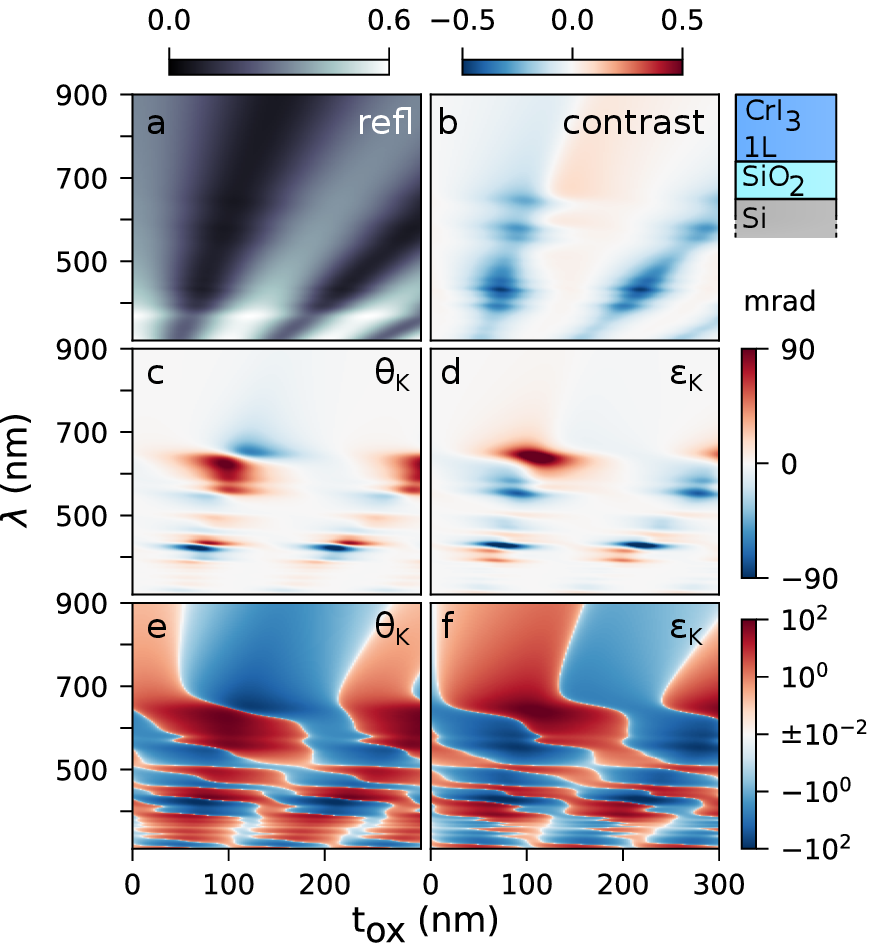}
    \caption{Simulation results for a CrI$_3$(1L)-SiO\ts{2}(285 nm)-Si heterostructure. The dielectric tensor of CrI$_3$ is taken from Molina-Sánchez et al. \cite{Molina-Sanchez2020} (Fig. \ref{fig:supp_eps_CrI3_ml_MolinaSanchez2020}). The reflectivity (a), contrast (b), Kerr rotation (c, e) and ellipticity (d, f) are shown as function of wavelength and oxide thickness. Note that the color scale used in (e) and (f) is a symmetric log scale that is cut at $\pm 10^{-2}$ mrad. All values between $\pm 10^{-2}$ mrad are indicated by the color white. Where the color scale is saturated, the values exceed the bounds of the scale.}
    \label{fig:supp_cri3-molina-sanchez_rot_vs_wl_sio2-thickness}
\end{figure*}

\begin{figure*}[htp]
    \centering
    \includegraphics[width=0.75\textwidth]{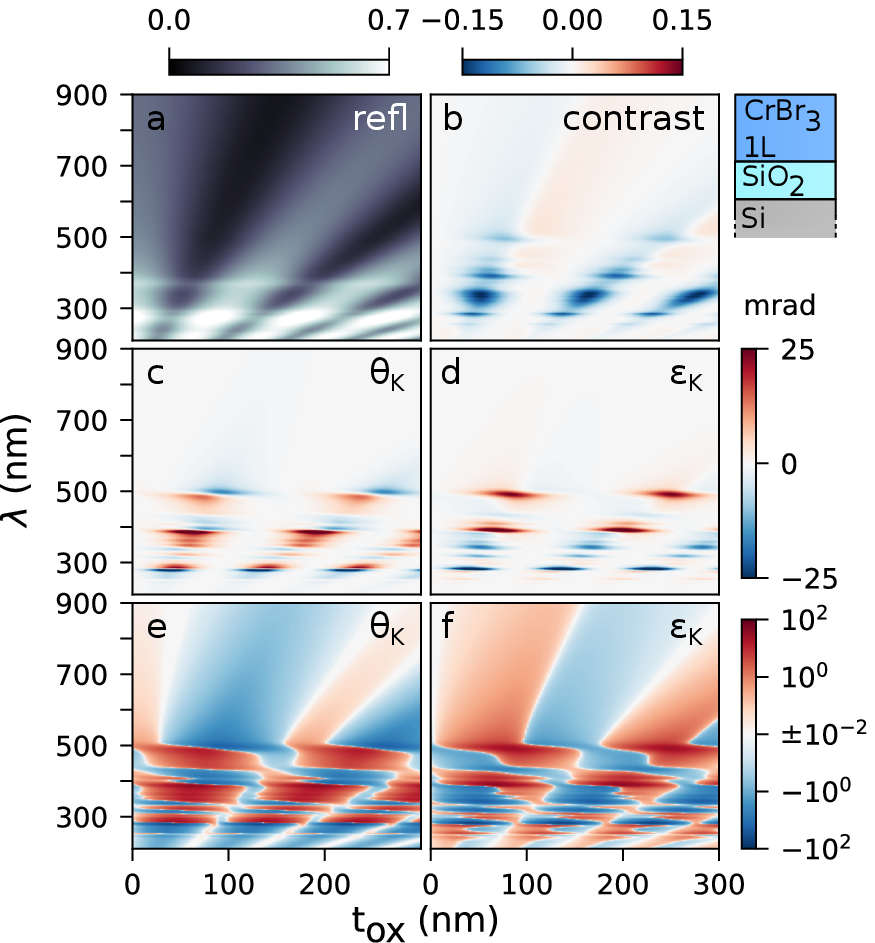}
    \caption{Simulation results for a CrBr$_3$(1L)-SiO\ts{2}(285 nm)-Si heterostructure. The dielectric tensor of CrBr$_3$ is taken from Molina-Sánchez et al. \cite{Molina-Sanchez2020} The reflectivity (a), contrast (b), Kerr rotation (c, e) and ellipticity (d, f) are shown as function of wavelength and oxide thickness. Note that the color scale used in (e) and (f) is a symmetric log scale that is cut at $\pm 10^{-2}$ mrad. All values between $\pm 10^{-2}$ mrad are indicated by the color white. Where the color scale is saturated, the values exceed the bounds of the scale.}
    \label{fig:supp_crbr3-molina-sanchez_rot_vs_wl_sio2-thickness}
\end{figure*}

\begin{figure*}[htp]
    \centering
    \includegraphics[width=0.75\textwidth]{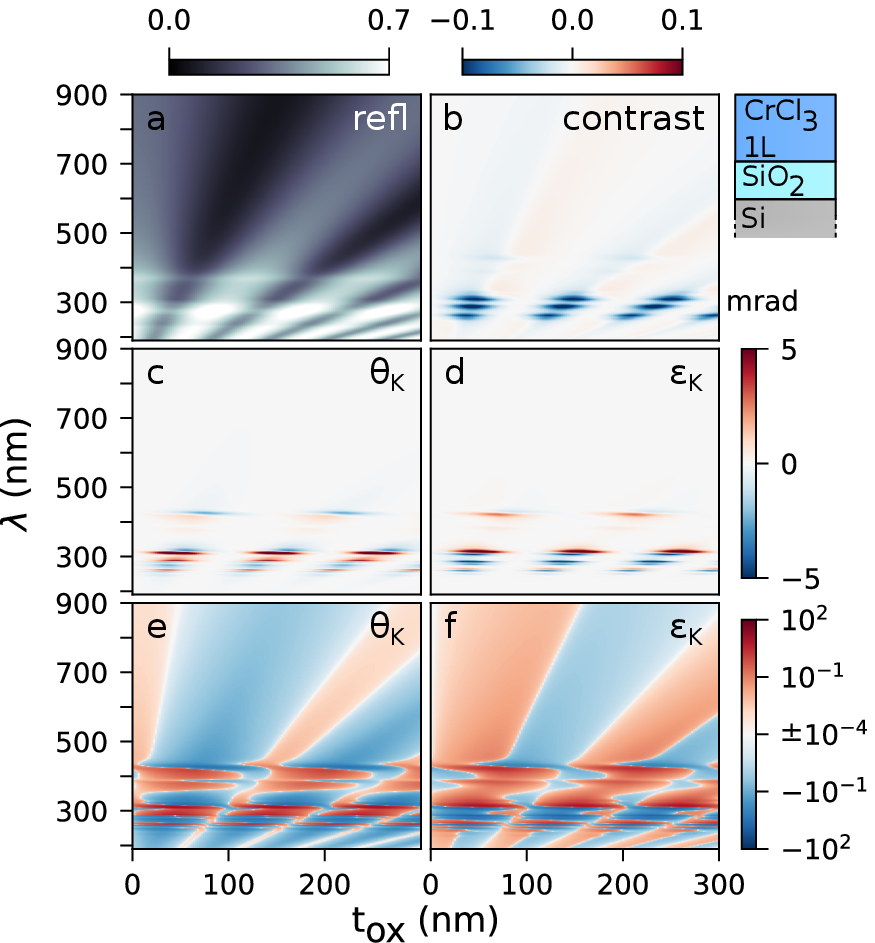}
    \caption{Simulation results for a CrCl$_3$(1L)-SiO\ts{2}(285 nm)-Si heterostructure. The dielectric tensor of CrCl$_3$ is taken from Molina-Sánchez et al. \cite{Molina-Sanchez2020} The reflectivity (a), contrast (b), Kerr rotation (c, e) and ellipticity (d, f) are shown as function of wavelength and oxide thickness. Note that the color scale used in (e) and (f) is a symmetric log scale that is cut at $\pm 10^{-4}$ mrad. All values between $\pm 10^{-4}$ mrad are indicated by the color white. Where the color scale is saturated, the values exceed the bounds of the scale.}
  \label{fig:supp_crcl3-molina-sanchez_rot_vs_wl_sio2-thickness}
\end{figure*}

\begin{figure*}[htp]
    \centering
    \includegraphics[width=0.75\linewidth]{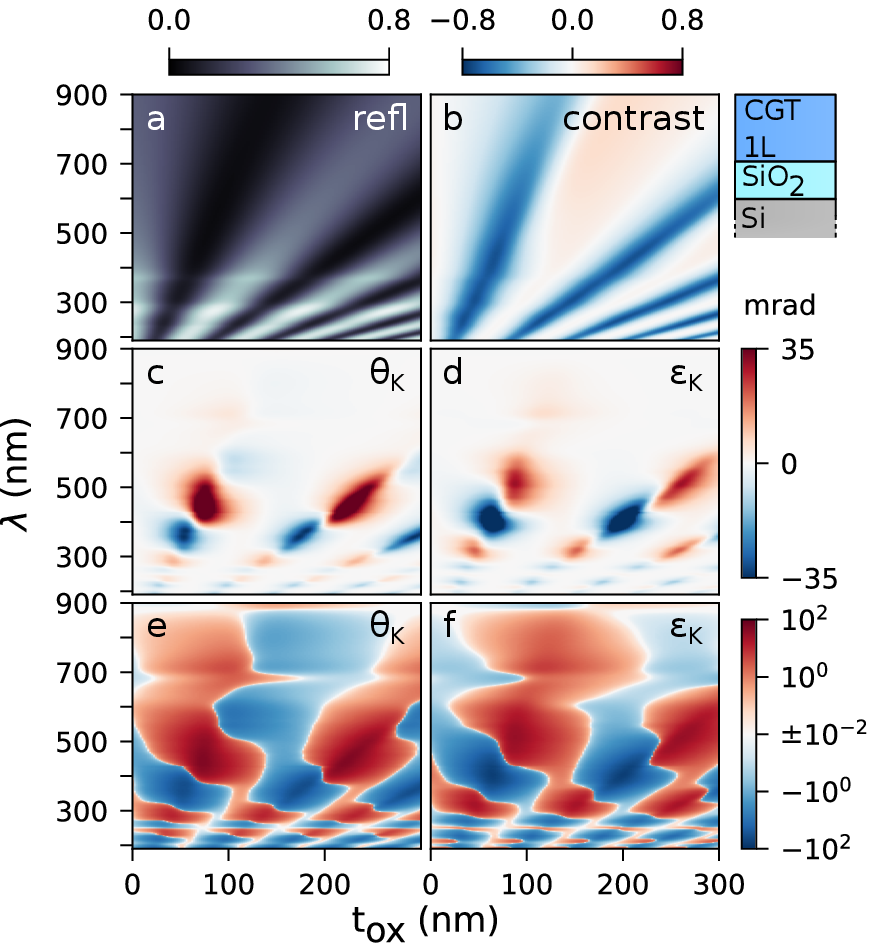}
    \caption{Simulation results for a Cr$_2$Ge$_2$Te$_6$(1L)-SiO\ts{2}(285 nm)-Si heterostructure. The dielectric tensor of Cr$_2$Ge$_2$Te$_6$ is taken from Fang et al. \cite{Fang2018} The reflectivity (a), contrast (b), Kerr rotation (c, e) and ellipticity (d, f) are shown as function of wavelength and oxide thickness. Note that the color scale used in (e) and (f) is a symmetric log scale that is cut at $\pm 10^{-2}$ mrad. All values between $\pm 10^{-2}$ mrad are indicated by the color white. Where the color scale is saturated, the values exceed the bounds of the scale.}
  \label{fig:supp_cgt_rot_vs_wl_sio2-thickness}
\end{figure*}

\FloatBarrier

\section{Bulk C\lowercase{r}I$_3$  with experimental values for the diagonal of its dielectric tensor}
\label{S_sec:exp_diag}
\begin{figure*}[htp]
    \centering
    \includegraphics[width=0.75\textwidth]{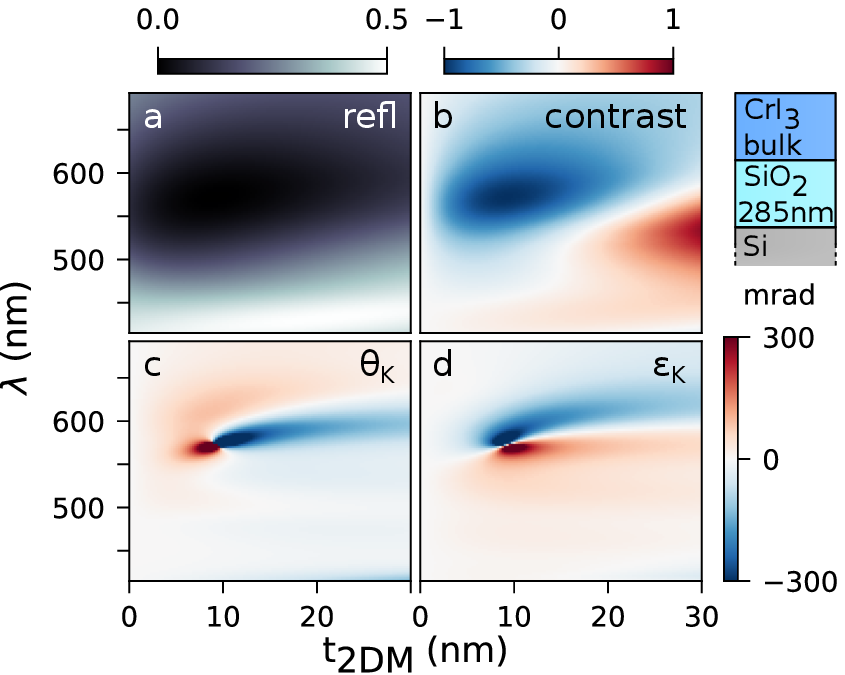}
    \caption{Simulation results for a stack of 
    CrI\ts{3}(bulk)-SiO\ts{2}(285 nm)-Si stack. Experimental values are used for the diagonal of the dielectric tensor of CrI\ts{3}, \cite{Grant1968, Huang2017} and theoretical vales for the off-diagonal elements. \cite{Wu2019} The reflectivity (a), contrast (b), Kerr rotation (c) and ellipticity (d) are shown as function of wavelength and oxide thickness. Where the color scale is saturated, the values exceed the bounds of the scale.}
    \label{fig:cri3-experimental-diag_rot_vs_wl_cri3-thickness}
\end{figure*}

For bulk CrI\ts{3}, experimental data is available for the diagonal elements of its dielectric tensor \cite{Grant1968, Huang2017}. Fig. \ref{fig:cri3-experimental-diag_rot_vs_wl_cri3-thickness} shows the simulation results for the same stack as in Fig. 3 of the main text, but here the experimental values are used for the diagonal of the dielectric tensor of CrI\ts{3}.
Using the experimental instead of theoretical values changes the wavelength and CrI\ts{3} thickness for which $\theta_K$ and $\varepsilon_K$ are maximized, and it slightly changes the shape of the patterns seen in the plots of the reflectivity, contrast, Kerr rotation and ellipticity.
However, the Kerr signals still have a similar non-monotonic behavior, showing a strong peak and a dip where the reflectivity is close to a minimum.

\FloatBarrier

\clearpage

\section{Signal to noise ratio}
To find the best signal to noise ratio (SNR), we take $R\Theta \overset{\mathrm{def}}{=} R\sqrt{\theta_K^2 + \varepsilon_K^2}$ as a measure for the signal strength, \cite{Challener1987} and consider two different noise regimes. In the case where the noise is independent of illumination intensity, the SNR is proportional to the signal strength (Fig. \ref{fig:cri3-experimental-diag_rot_vs_wl_cri3-thickness}e). If the noise is dominated by shot noise, which grows as $\sqrt{I}$, the SNR is proportional to $R\Theta / \sqrt{R} = \sqrt{R}\Theta$, which is shown in Fig. \ref{fig:cri3-experimental-diag_rot_vs_wl_cri3-thickness}f. In both cases, the highest SNR is attained at specific wavelength intervals. In the shot noise dominated regime, the best SNR is obtained when the MOKE signals are (almost) maximized, while for the intensity independent noise regime, the best SNR is obtained when the MOKE signals are further away from their maximum, where the reflectivity is higher.
\begin{figure*}[htp]
    \centering
    \includegraphics[width=0.75\textwidth]{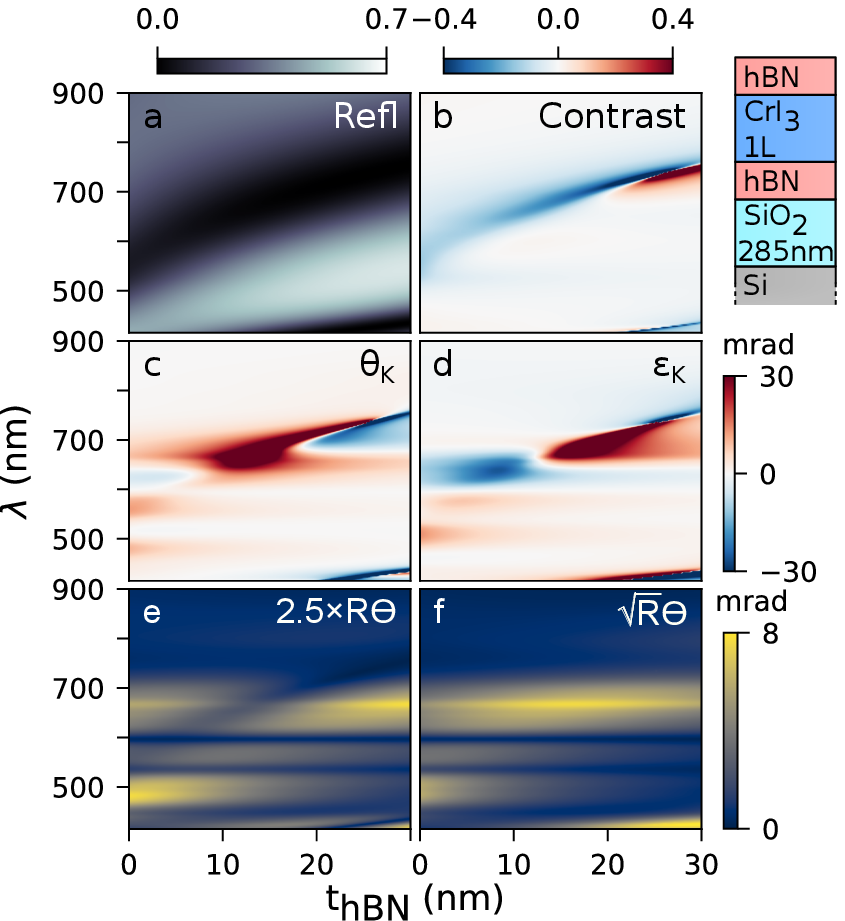}
    \caption{Simulation results for a hBN-CrI$_3$(1L)-hBN-SiO$_2$-Si(285nm) stack, including an estimation for the signal to noise ratio (SNR) in two different noise regimes, as a function of wavelength and hBN thickness. The reflectivity (a), contrast (b), Kerr rotation (c) and ellipticity (d) are also shown in the main text in Fig. 4. The signal strength $R\Theta$ shown in (e) is proportional to the SNR for the case where the noise is independent of the intensity of the light. If the noise is dominated by shot noise, the SNR is proportional to $\sqrt{R}\Theta$, shown in (f). Where the color scale is saturated, the values exceed the bounds of the scale.}
    \label{fig:snr_hbn-cri3-hbn-sio2-si_vs_wl_hbn-thickness}
\end{figure*}

\clearpage
\section{Constant dielectric tensor for 2D magnets}
\begin{figure*}[htp]
    \centering
    \includegraphics[width=0.75\textwidth]{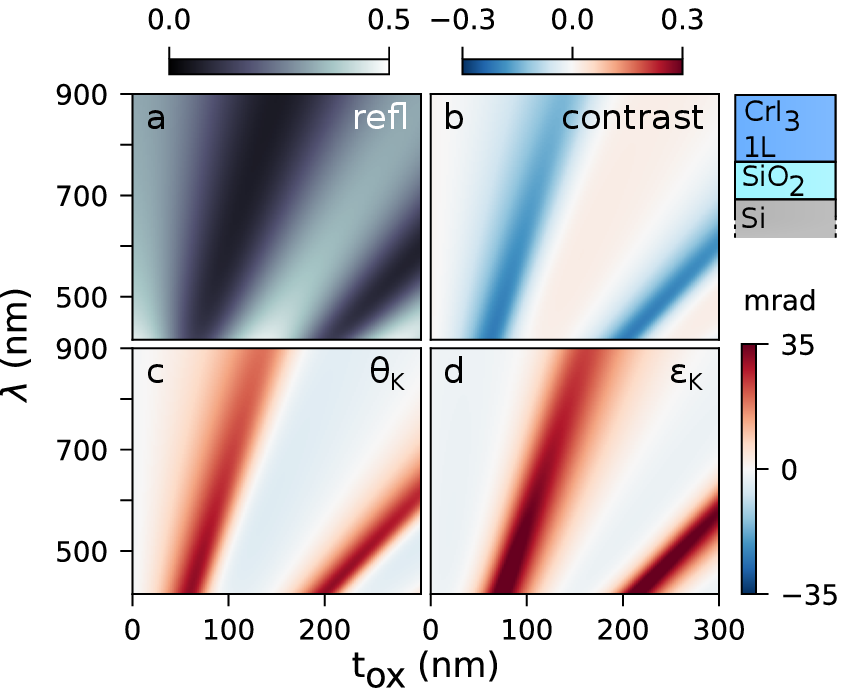}
    \caption{Simulation results for a CrI\ts{3}(1L)-SiO\ts{2}(285 nm)-Si heterostructure where the dielectric tensor of CrI\ts{3} is fixed to is value at 680 nm. The reflectivity (a), contrast (b), Kerr rotation (c) and ellipticity (d) are shown as function of wavelength and CrI\ts{3} thickness. Where the color scale is saturated, the values exceed the bounds of the scale.}
    \label{fig:cri3-wl680nm-_rot_vs_wl_sio2-thickness}
\end{figure*}

To determine the effect of the wavelength dependence of the dielectic tensor of CrI\ts{3} on our results, we perform the same simulation as done for Fig. 2 in the main text, but with a dielectric tensor of CrI\ts{3} that is fixed to its value at a wavelength of 680 nm.
The results displayed in Fig. \ref{fig:cri3-wl680nm-_rot_vs_wl_sio2-thickness}, unlike the results for the wavelength-dependent dielectric tensor of CrI\ts{3} displayed in Fig. 2 of the main text, do not show any additional features on top of the fan pattern, which is the dominant pattern for the reflectivity.
This indicates that the additional features seen in the figures of the main text and in sections \ref{S_sec:other_magnets} and \ref{S_sec:exp_diag}, are caused by the wavelength dependence of the 2D magnet.

\FloatBarrier
\clearpage
\section{Large MOKE signals at low reflectivity}
A common feature in all our simulations is that the magnitudes of $\theta_K$ and $\varepsilon_K$ are maximized when the reflectivity is close to a minimum.
This can be explained by the behaviour of the reflection coefficients for the electric field of the two circular polarizations, $r_+$ and $r_-$.
Since the circular polarizations are eigenmodes for all materials we used, reflection off the heterostructure does not mix them in the polar geometry.
Therefore the two circular polarizations can be treated independently.

\begin{figure*}[htp]
    \centering
    \includegraphics[width=\textwidth]{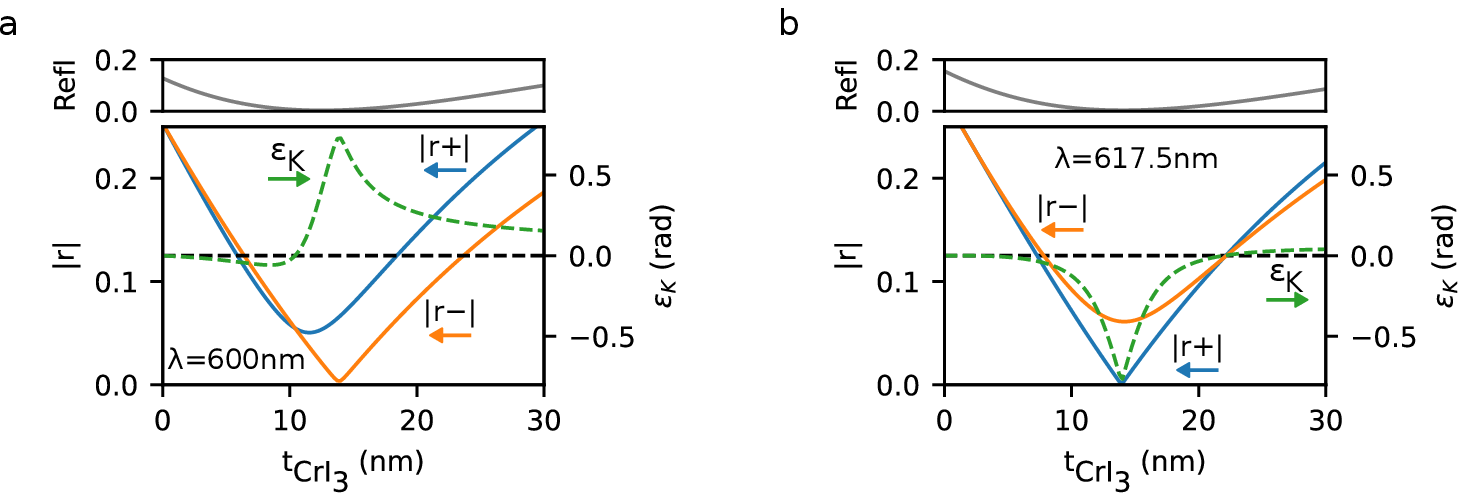}
    \caption{Simulated $|r_+|$, $|r_-|$ and $\varepsilon_K$ for a CrI\ts{3}(bulk)-SiO\ts{2}(285nm)-Si stack, plotted as a function of CrI\ts{3} thickness. Results are shown for a wavelength of 600 nm (a), and 617.5 nm (b).}
    \label{fig:large_ellipticity_vs_thickness}
\end{figure*}

\begin{figure*}[htp]
    \centering
    \includegraphics[width=\textwidth]{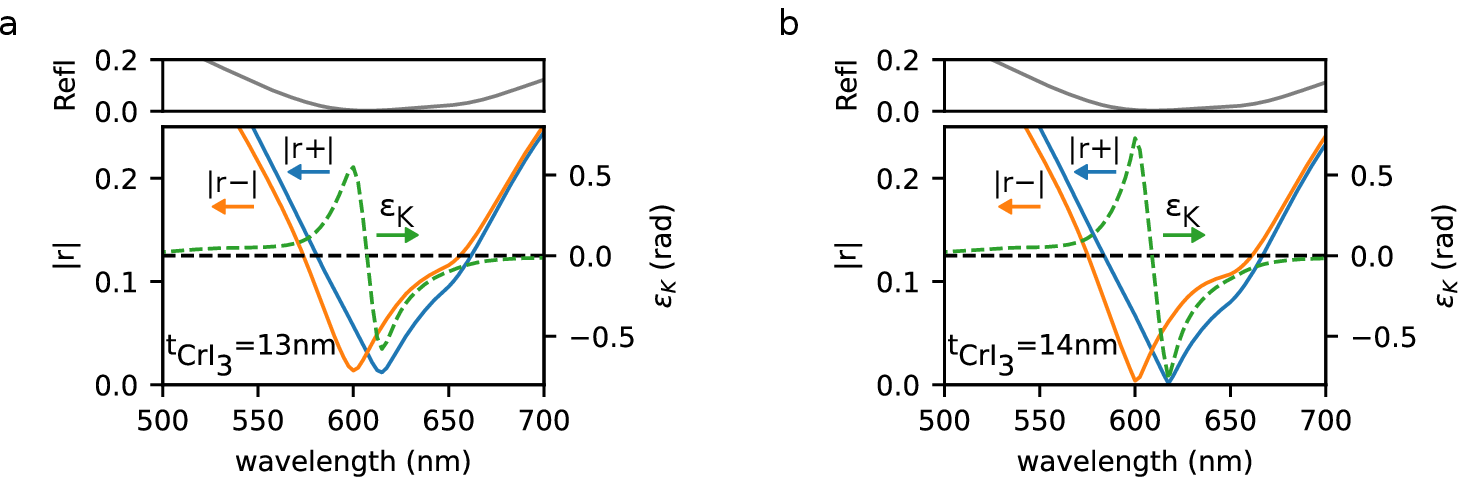}
    \caption{Simulated $|r_+|$, $|r_-|$ and $\varepsilon_K$ for a CrI\ts{3}(bulk)-SiO\ts{2}(285nm)-Si stack, plotted as a function of wavelength. Results are shown for a CrI\ts{3} thickness of 13 nm (a), and 14nm (b).}
    \label{fig:large_ellipticity_vs_wavelength}
\end{figure*}

Due to the small circular birefringence and dichroism induced by the 2DM, the reflectivity for the two circular polarizations are only slightly different.
Therefore, when the reflectivity (defined as $|r_+|^2 + |r_-|^2$) has a minimum, the reflectivity of both circular polarizations must be very close to their minimum.
This means that the ellipticity, defined as $\tan^{-1}(|r_+| - |r_-|) / (|r_+| + |r_-|)$, has its extrema close to this reflectivity minimum.
In Fig. \ref{fig:large_ellipticity_vs_thickness} and \ref{fig:large_ellipticity_vs_wavelength}, $|r_+|$, $|r_-|$ and $\varepsilon_K$ are plotted for the stack used in Fig. 3 of the main text, which consists of bulk CrI\ts{3} on top of a SiO\ts{2}(285nm)/Si substrate.
The former shows it as a function of CrI\ts{3} thickness at a constant wavelength, and the latter as a function of wavelength at a constant CrI\ts{3} thickness.
These figures indicate that the minima of $r_+$ and $r_-$ are indeed close to the reflectivity minimum, and that their minima have in general different minimum values and are located at different positions.

\begin{figure*}[htp]
    \centering
    \includegraphics[width=\textwidth]{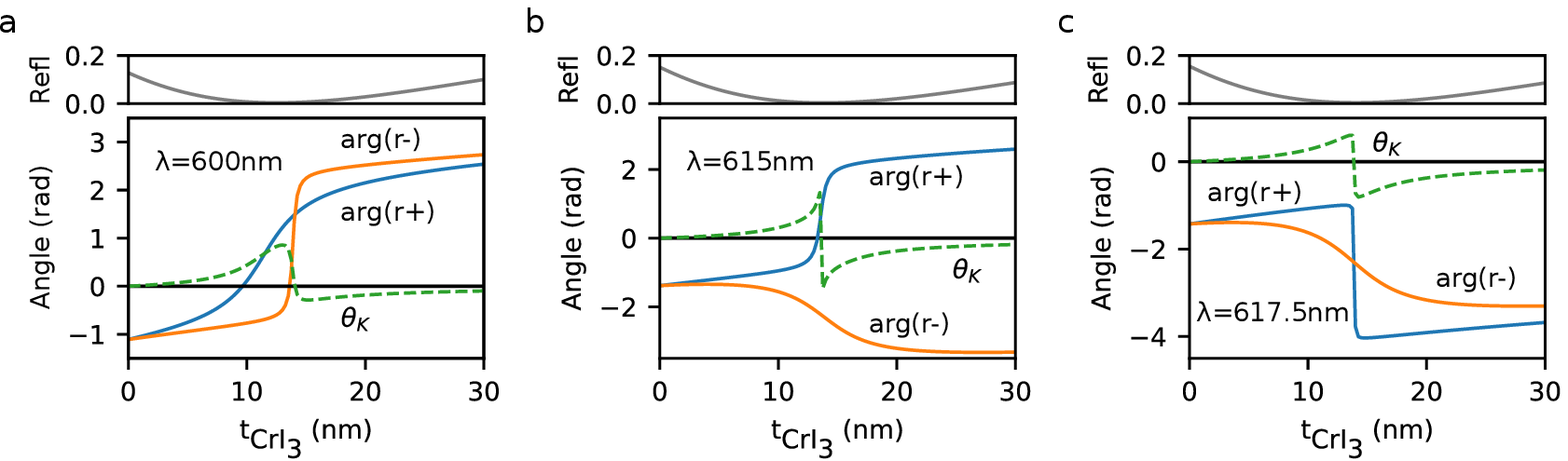}
    \caption{Simulated $\text{arg}(r_+)$, $\text{arg}(r_-)$ and $\theta_K$ for a CrI\ts{3}(bulk)-SiO\ts{2}(285nm)-Si heterostructure, plotted as a function of CrI\ts{3} thickness. Results are shown for a wavelength of (a) 600 nm, (b) 615 nm, and (c) 617.5 nm.}
    \label{fig:large_rotation_vs_thickness}
\end{figure*}

\begin{figure*}[htp]
    \centering
    \includegraphics[width=\textwidth]{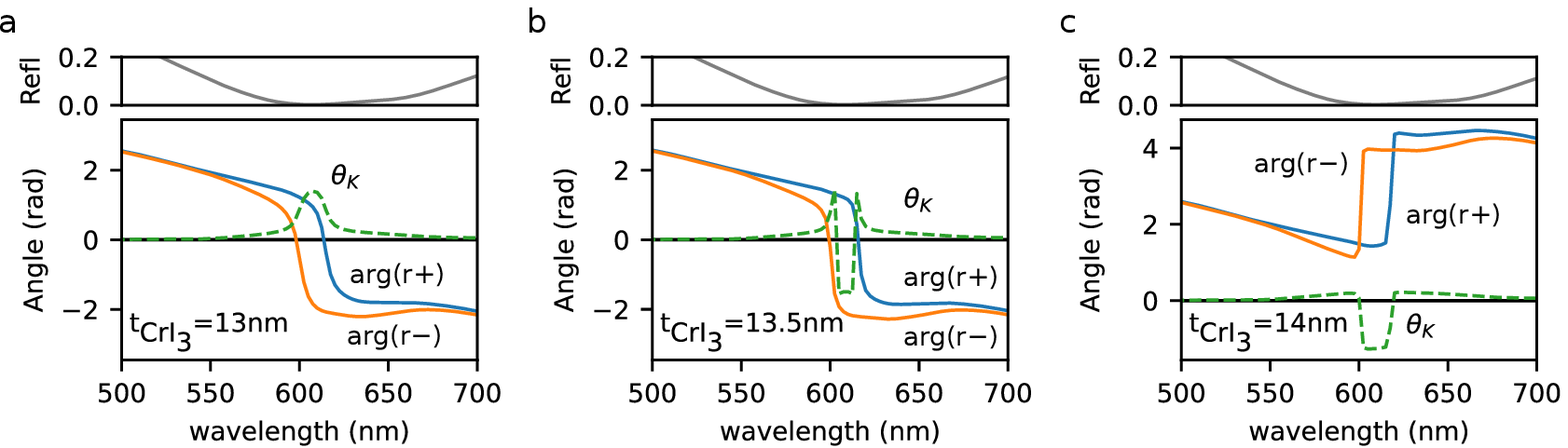}
    \caption{Simulated $\text{arg}(r_+)$, $\text{arg}(r_-)$ and $\theta_K$ for a CrI\ts{3}(bulk)-SiO\ts{2}(285nm)-Si heterostructure, plotted as a function of wavelength. Results are shown for a CrI\ts{3} thickness of (a) 13 nm, (b) 13.5 nm, and (c) 14 nm.}
    \label{fig:large_rotation_vs_wavelength}
\end{figure*}

The large Kerr rotations can be explained by the behavior of the complex phase of $r_\pm$ when they are approximated to be linear in wavelength and layer thickness close to the reflectivity minimum.
Let $r_\pm(x) = r_\pm(x_0) + r'_\pm(x_0) \Delta x$, where $x$ is the wavelength, thickness, or a linear combination of the two, $r'_\pm(x)$ is the derivative of $r_\pm(x)$ with respect to $x$, $\Delta x$ is $x - x_0$, and $x_0$ is the position of the reflectivity minimum.
Around $x_0$, the phase of the complex reflectivity coefficients changes by $\pi$ (from $\text{arg}(-r'_\pm \Delta x)$ to $\text{arg}(r'_\pm \Delta x)$) at a scale of $|r_\pm(x_0) \sin(\psi)|$, where $\psi$ is the difference in complex phase of $r_\pm(0)$ and $r'_\pm$.
The change in phase of $r_+$ and $r_-$ are in general centered at slightly different values of $x$, and happen at different scales, because the ferromagnetic layer has a different refractive index for the two circular polarizations.
Therefore, the Kerr rotation, given by $\theta_K = (\text{arg}(r_+) - \text{arg}(r_-))/2$, increases when $|r_\pm(x_0) \sin(\psi)|$ becomes smaller.
This happens when $|r_\pm(0)|$ is small, i.e. when the reflectivity is close to its minimum.
In Fig. \ref{fig:large_rotation_vs_thickness} and \ref{fig:large_rotation_vs_wavelength}, $\text{arg}(r_+)$, $\text{arg}(r_-)$ and $\theta_K$ are plotted for the stack used in Fig. 3 of the main text, which consists of bulk CrI\ts{3} on top of a SiO\ts{2}(285nm)/Si substrate.
The former shows it as a function of CrI\ts{3} thickness at a constant wavelength, and the latter as a function of wavelength at a constant CrI\ts{3} thickness.
These figures indicate that $\text{arg}(r_+)$ and $\text{arg}(r_-)$ indeed change rapidly close to the reflectivity minimum, and that this happens in general at different scales and at different positions.
The Kerr rotation is mapped to the interval $(-\pi, \pi]$ .
When the phase difference of $r_+$ and $r_-$ crosses $\pm\pi$, the Kerr rotation is mapped back into this interval, giving rise to discontinuities in $\theta_K$.

\FloatBarrier

\bibliographystyle{aapmrev4-1}
\bibliography{supplement.bib}